\documentclass[sigconf,nonacm]{acmart}

\AtBeginDocument{%
  }

\renewcommand\footnotetextcopyrightpermission[1]{} 

\usepackage{enumitem}
\usepackage[table]{xcolor}    
\usepackage{booktabs}         
\usepackage{graphicx}         
\usepackage{pifont}           
\usepackage{array}            

\usepackage{arydshln}
\usepackage{caption}
\usepackage{caption}
\usepackage{array,booktabs,makecell}
\usepackage{float}
\usepackage{tabularx} 
\usepackage{makecell}

\newcolumntype{C}{>{\centering\arraybackslash}X}

\usepackage{xcolor}
\definecolor{lightgraytext}{RGB}{140,140,140}

\newcommand{\pair}[2]{#1{\color{lightgraytext}/#2}}
\usepackage{booktabs} 
\usepackage{tabularx} 
\usepackage{multirow} 
\usepackage{xcolor}   
\usepackage{makecell}
\usepackage{caption}
\usepackage{dblfloatfix}
\usepackage{booktabs}   
\usepackage{tabularx}   
\usepackage{multirow}   
\usepackage{placeins}
\usepackage{balance}
\usepackage{cuted}

\begin{document}

\title{VidAudio-Bench: Benchmarking V2A and VT2A Generation across Four Audio Categories}

\author{Qian Zhang}
\affiliation{%
  \institution{Shanghai Jiaotong University}
  \state{Shanghai}
  \country{China}
}
\email{zq1729@sjtu.edu.cn}

\author{Yuqin Cao}
\affiliation{%
  \institution{Shanghai Jiaotong University}
  \state{Shanghai}
  \country{China}}
\email{caoyuqin@sjtu.edu.cn}

\author{Yixuan Gao}
\affiliation{%
  \institution{Shanghai Jiaotong University}
  \state{Shanghai}
  \country{China}
}
\email{gaoyixuan@sjtu.edu.cn}

\author{Xiongkuo Min}
\affiliation{%
  \institution{Shanghai Jiaotong University}
  \state{Shanghai}
  \country{China}}
\email{minxiongkuo@sjtu.edu.cn}

\renewcommand{\shortauthors}{Zhang et al.}

\begin{abstract}
Video-to-Audio (V2A) generation is essential for immersive multimedia experiences, yet its evaluation remains underexplored. Existing benchmarks typically assess diverse audio types under a unified protocol, overlooking the fine-grained requirements of distinct audio categories. To address this gap, we propose \textbf{VidAudio-Bench}, a multi-task benchmark for V2A evaluation with four key features: (1) \textbf{Broad Coverage}: It encompasses four representative audio categories—sound effects, music, speech, and singing—under both V2A and Video-Text-to-Audio (VT2A) settings. (2) \textbf{Extensive Evaluation}: It comprises 1,634 video-text pairs and benchmarks 11 state-of-the-art generation models. (3) \textbf{Comprehensive Metrics}: It introduces 13 task-specific, reference-free metrics to systematically assess audio quality, video–audio consistency, and text–audio consistency. (4) \textbf{Human Alignment}: It validates all metrics through subjective studies, demonstrating strong consistency with human preferences. Experimental results reveal that current V2A models perform poorly in speech and singing compared to sound effects. Our VT2A results further highlight a fundamental tension between instruction following and visually grounded generation: stronger visual conditioning improves video-audio alignment, but often at the cost of generating the intended audio category. These findings establish VidAudio-Bench as a comprehensive and scalable framework for diagnosing V2A systems and provide new insights into multimodal audio generation. 
\end{abstract}


\begin{teaserfigure}
  \includegraphics[width=\textwidth]{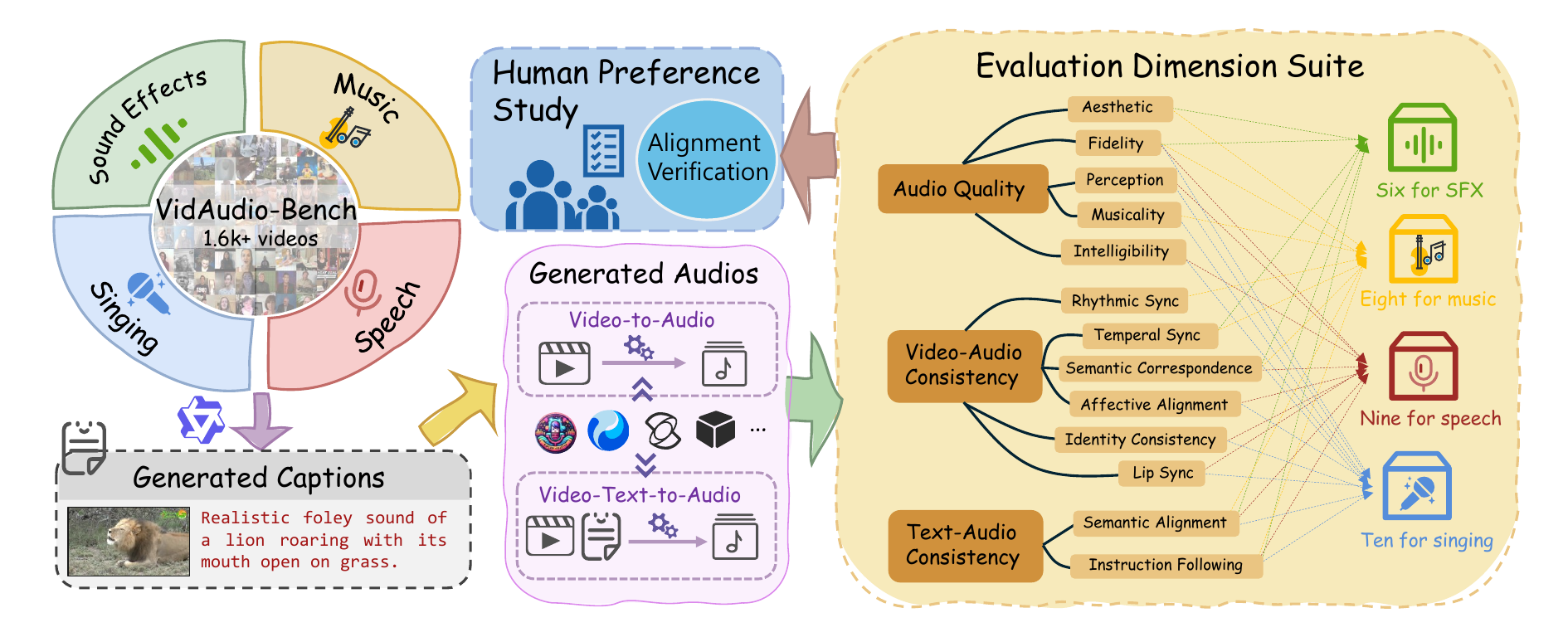}
  \caption{\textbf{Overview of VidAudio-Bench.} We categorize audio generation into four task types: \textit{sound effects}, \textit{music}, \textit{speech}, and \textit{singing}. We further introduce two input paradigms, V2A and VT2A, to analyze how adding textual descriptions changes generation behavior. Our evaluation suite spans Audio Quality, Video-Audio Consistency, and Text-Audio Consistency, covering 13 fine-grained dimensions. Human preference studies show strong correlation between our metrics and human perception.}
  \label{fig:teaser}
\end{teaserfigure}

\maketitle
\thispagestyle{plain}
\pagestyle{plain}
\section{Introduction}
While audio generation has made significant progress through Text-to-Audio (T2A)~\cite{kreukaudiogen, yang2023diffsound, yang2021multi, huang2023make, melechovsky2024mustango, zivmasked} and Image-to-Audio (I2A)~\cite{chen2024images, sheffer2023hear, iashin2021taming} models, relying solely on static or textual inputs often fails to capture precise temporal dynamics, complex spatial environments, and realistic physical interactions. This limitation has driven the shift toward Video-to-Audio (V2A) generation~\cite{luo2023diff, xing2024seeing, wang2024v2a, jeong2025read, xie2024sonicvisionlm, zhang2026foleycrafter}, where rich visual cues provide explicit conditions for dynamic audio synthesis. 
Recent V2A models have evolved toward multimodal and multi-task frameworks~\cite{wang2025kling, xu2025uniflow, dai2026omni2sound}. Pioneering models such as AudioX~\cite{tian2025audiox} and AudioGen-Omni~\cite{wang2025audiogen} enable the flexible generation of diverse audio types, emphasizing the importance of not just sound effects, but also music and speech. Similarly, large-scale frameworks like Kling-Foley~\cite{wang2025kling} and Audiobox-Aesthetics~\cite{tjandra2025meta} categorize their training data into distinct modalities (\textit{e.g.}, sound effects, music, speech and singing). These advances call for more fine-grained evaluation protocols that can reflect the specific requirements of different audio categories.

However, existing evaluation methodologies remain monolithic and outdated.
First, current benchmarks suffer from the "one-size-fits-all" pitfall. Despite the distinct acoustic and semantic properties of different audio tasks, generative models are almost exclusively evaluated using generic distribution-level metrics, such as Fréchet Audio Distance (FAD)~\cite{roblek2019fr}, Kullback-Leibler Divergence (KL), Inception Score (IS)~\cite{salimans2016improved}, and cross-modal similarity (\textit{e.g.}, ImageBind~\cite{girdhar2023imagebind}). These metrics fail to capture task-dependent requirements, such as the intelligibility and lip-sync precision necessary for speech, or the melodic coherence required for music. Second, the evaluation data itself is problematic. Most models still evaluate on the raw VGG-Sound test set~\cite{chen2020vggsound}, which, as explicitly noted by the VGGSounder~\cite{zverev2025vggsounder}, suffers from severe limitations including co-occurring classes, overlapping sounds, and modality misalignment.

A further critical bottleneck in current V2A research lies in the ambiguous role of text. Early models such as FRIEREN~\cite{wang2024frieren} rely solely on visual inputs, whereas most recent architectures naturally support joint video–text conditioning (VT2A)~\cite{wang2025kling, xu2025uniflow, shan2025hunyuanvideo, tian2025audiox, liuthinksound}. However, during evaluation, it remains unclear whether the generated audio is successfully grounded in the visual content, or if the model mainly follows explicit acoustic textual prompts (\textit{e.g.}, "\textit{the sound of a dog barking}") as a shortcut. This calls for an evaluation paradigm that can disentangle models' visual understanding capability from auxiliary text guidance, which is particularly vital for context-aware multimedia applications.

To address these limitations, we propose \textbf{VidAudio-Bench}, the first comprehensive multi-task benchmark designed for both V2A and VT2A evaluation. As shown in Figure~\ref{fig:teaser}, VidAudio-Bench encompasses four representative audio categories: sound effects (SFX), music, speech, and singing. For each category, we construct carefully curated evaluation subsets, totaling 1,634 video-text pairs with strong audio-visual correlation. 
To move beyond a monolithic evaluation, we introduce a suite of task-specific, reference-free evaluation protocols. VidAudio-Bench comprises 13 fine-grained dimensions covering audio quality, cross-modal alignment, and domain-specific attributes. Leveraging recent advances in multimodal large language models (MLLMs)~\cite{comanici2025gemini, hurst2024gpt, sun2024video, xu2025qwen3, zhang2023video}, which have been shown to be effective for evaluation~\cite{liang2026omni}, we incorporate an MLLM-as-a-Judge framework for three of our advanced dimensions. Crucially, comprehensive user studies confirm that our evaluation suite is closely aligned with human perception.

Moreover, VidAudio-Bench introduces a novel VT2A evaluation paradigm that explicitly probes a model’s visual understanding. Instead of providing textual prompts that dictate the target sound, we supply dense visual descriptions of the scene. This zero-information-leak design prevents models from exploiting acoustic textual shortcuts, enabling a cleaner test of visual grounding and instruction following. Our results further reveal a counterintuitive effect: dense captions often improve semantic alignment but weaken instruction following, leading to target-miss errors.

In summary, our main contributions are threefold:
\begin{itemize}[leftmargin=*,nosep]
    \item \textbf{Comprehensive Multi-Task Benchmark:} We present VidAudio-Bench, which organizes video-to-audio generation into four sub-tasks: SFX, music, speech, and singing. Featuring over 400 highly correlated video-text pairs per task and 13 fine-grained dimensions, it provides a more systematic evaluation framework.
    \item \textbf{Novel VT2A Evaluation Setting:} We introduce a VT2A evaluation setting using dense visual descriptions instead of explicit audio prompts, enabling a cleaner assessment of visual understanding and reducing shortcut reliance on textual acoustic cues.
    \item \textbf{Extensive Benchmarking and Insights:} 
    We benchmark a broad range of state-of-the-art models and show that current systems still struggle with domain-specific generation and visually grounded audio synthesis. Our study further uncovers the dual role of visual prompts in VT2A generation.
\end{itemize}

\section{Related Works}
\subsection{Audio Generation Models}
Video-to-Audio (V2A) aims to generate semantically aligned and temporally synchronized audio for silent videos. 
Early methods~\cite{zhou2018visual} directly mapped frames to waveforms, while later methods like SpecVQGAN~\cite{iashin2021taming} and Im2Wav~\cite{sheffer2023hear} generate latent audio representations conditioned on visual features extracted by models like CLIP~\cite{radford2021learning}. 
To overcome data scarcity, recent work leverages pretrained T2A models~\cite{ zhang2026foleycrafter, wang2024v2a, xie2024sonicvisionlm}. For example, Seeing and Hearing~\cite{xing2024seeing} converts videos into AudioLDM~\cite{liu2023audioldm} prompts via ImageBind~\cite{girdhar2023imagebind}, while V2A-Mapper~\cite{wang2024v2a} aligns visual features with CLAP~\cite{wu2023large}. 
To further improve temporal alignment, Diff-Foley~\cite{luo2023diff} introduces contrastive audio-visual pretraining, whereas FoleyCrafter~\cite{zhang2026foleycrafter}, ReWaS~\cite{jeong2025read}, and SonicVLM~\cite{xie2024sonicvisionlm} integrate time-aware control modules.
Recent efforts also focus on alignment and efficiency. V-AURA~\cite{viertola2025temporally} uses high-frame-rate visual features, and FRIEREN~\cite{wang2024frieren} accelerates sampling via efficient rectified flow matching (RFM). 
Furthermore, unified multimodal training (\textit{e.g.}, VATT~\cite{akbari2021vatt}, MMAudio~\cite{cheng2025mmaudio}) has been explored to improve semantic consistency.
Beyond this, some works investigate more flexible control. MultiFoley~\cite{chen2025video} supports multimodal conditioning, and ThinkSound~\cite{liuthinksound} introduces Chain-of-Thought (CoT) reasoning for guided synthesis. 
More recent models, such as HunyuanVideo-Foley~\cite{shan2025hunyuanvideo} and Kling-Foley~\cite{wang2025kling}, adopt advanced diffusion transformer architectures to further improve audio quality and synchronization.

\subsection{Evaluation Benchmarks}
The rapid progress of Artificial Intelligence Generated Content (AIGC) has driven the development of systematic evaluation benchmarks. Existing frameworks such as VBench~\cite{huang2024vbench} and EvalCrafter~\cite{liu2024evalcrafter} provide multi-dimensional evaluation for text-to-video (T2V) generation, while TTA-Bench~\cite{wang2026tta} and T2A-EpicBench~\cite{wang2025t2a} focus on T2A generation quality.
More recently, benchmarks for joint audio-video generation have also emerged. For example, VABench~\cite{hua2025vabench} presents a 15-dimension framework for evaluating Text-to-Audio-Video (T2AV) and Image-to-Audio-Video (I2AV), and T2AV-Compass~\cite{cao2025t2av} combines objective signal-level metrics with subjective MLLM-as-a-Judge evaluation.
However, existing benchmarks do not fully address the challenges of V2A evaluation.
Unlike T2AV, V2A requires models to infer plausible audio content from visual cues alone, leading to greater ambiguity and a stronger need for task-aware assessment. In addition, although T2AV-Compass groups audio into sounds, speech, and music, its evaluation remains relatively unified, without incorporating the distinct criteria required by different audio categories. Consequently, V2A evaluation is still underdeveloped and fragmented, and the field lacks a standardized multi-task benchmark. To this end, we propose VidAudio-Bench, a task-specific and reference-free benchmark for evaluating V2A and VT2A systems across diverse audio categories.
\vspace{1pt}
\section{Benchmark Construction}
\subsection{Task Design}
\label{subsec:task-design}
Motivated by recent advances in audio generation models~\cite{wang2025kling,wang2025audiogen}, we divide audio generation into four representative sub-tasks based on acoustic properties, semantic functions, and cross-modal alignment requirements. This taxonomy provides a structured framework for assessing the challenges of each audio category.

\noindent\textbf{Sound Effects (SFX).}
Sound effects (\textit{e.g.}, ambient, Foley, and interaction sounds) are tightly coupled with visual events. The key challenge lies in accurately recognizing the occurring events and generating sounds that are accurately synchronized with them.

\noindent\textbf{Music.}
Unlike transient sound effects, music is sustained and structured over time. Depending on the visual context, this task involves two distinct challenges, leading us to define two sub-categories:
\begin{itemize}[leftmargin=*,nosep]
\item {\textit{Instrumental Performance}}: Videos depicting musicians playing instruments. The generation must exhibit a frame-level alignment between visual actions (\textit{e.g.}, pressing piano keys, bowing a violin) and the resulting musical notes and rhythms.
\item {\textit{Background Music (BGM)}}: 
Videos requiring music to support the narrative atmosphere and emotional progression. Here, the focus shifts from strict temporal synchronization to broader semantic and affective alignment with the scene’s mood and pacing.
\end{itemize}

\noindent\textbf{Speech.}
The speech generation task focuses on synthesizing natural and high-fidelity human voices from talking faces. The key challenge is to produce intelligible speech that is precisely synchronized with lip movements and consistent with the speaker’s visible characteristics, such as timbre, age, gender, and expression.

\noindent\textbf{Singing.}
Singing combines characteristics of both speech and music. The main challenge is to generate melodious vocals that align with both the musical rhythm and the singer’s lip movements, while preserving lyric intelligibility and vocal identity consistency.
\subsection{Dataset Construction}
Corresponding to the four task definitions in Section~\ref{subsec:task-design}, we construct customized evaluation subsets for each task. This section details the selection criteria, source datasets, and dataset statistics.
\enlargethispage{\baselineskip}
\subsubsection{Data Selection Criteria}
A fundamental prerequisite for evaluating V2A generation is ensuring a \textit{high audio-visual correlation}. Specifically, the visual content should provide sufficient information to reliably predict the associated sounds. To ensure this, we establish strict filtering criteria for each task:

\begin{itemize}[leftmargin=2em, nosep]
    \item \textbf{Sound Effects:} Videos must contain a visible sound source and salient motion. The sound-producing objects must be strictly on-screen, explicitly excluding any off-screen voiceovers or ambient noises without visual grounding.
    \item \textbf{Music:} For \textit{Instrumental Performance}, both the musician and instrument must be clearly visible without severe occlusion or ambiguous visual cues. For \textit{Background Music}, the video must exhibit a distinct emotional mood or rhythmic cuts that naturally align with the musical narrative.
    \item \textbf{Speech and Singing:} Videos must present a single frontal face without occlusion. The video should clearly reveal lip movements; for singing samples, it should further provide evident expressive cues, such as facial expressions or upper-body motions, to facilitate reliable evaluation.
\end{itemize}
\vspace{-2pt}
\subsubsection{Source Datasets and Subset Construction}

To satisfy the requirements of the defined tasks, we curate suitable clips from several large-scale, high-quality public datasets. Further details on the subsets are provided in Appendix A.1.
\begin{itemize}[leftmargin=2em,nosep]
    \item \textbf{VGGSounder}~\cite{zverev2025vggsounder}: A re-annotated multi-label dataset containing 15,446 clips across 309 classes with over 40,000 labels. Its strong audio-visual grounding makes it suitable for event-centric audio generation. Based on our task taxonomy, we categorize its classes and sample 400 videos for the \textit{SFX} task and 191 videos for the \textit{Instrumental Performance} subset.
    \item \textbf{HarmonySet}~\cite{zhou2025harmonyset}: A large-scale video-music dataset containing 48,328 video-music pairs, in which background music is intentionally matched to the visual narrative. From this dataset, we extract 231 high-quality clips for the \textit{BGM} task.
    \item \textbf{AVSpeech}~\cite{ephrat2018looking}: A large-scale dataset containing 4,700 hours of clean, single-speaker lectures and TED talks, ensuring clear correspondence between speech and visible faces. From its test set, we sample 412 clips for the \textit{Speech} task.
    \item \textbf{Acappella}~\cite{montesinos2021cappella}: Designed for multimodal singing voice separation, this dataset comprises 46 hours of high-quality solo singing videos. We select 400 single-person English singing clips with unoccluded frontal faces for the \textit{Singing} task.
\end{itemize}
\vspace{-2pt}
\subsubsection{Data Processing and Statistics}
To accommodate the typical 10-second output window of current models, we standardized all clips to a uniform duration via precise trimming or padding. We retain only videos with a resolution of at least 720P to ensure high-quality visual input. In addition, all clips are stripped of their original audio so that models must rely solely on visual cues.

\begin{figure}[h]
  \centering

  \includegraphics[width=0.85\linewidth]{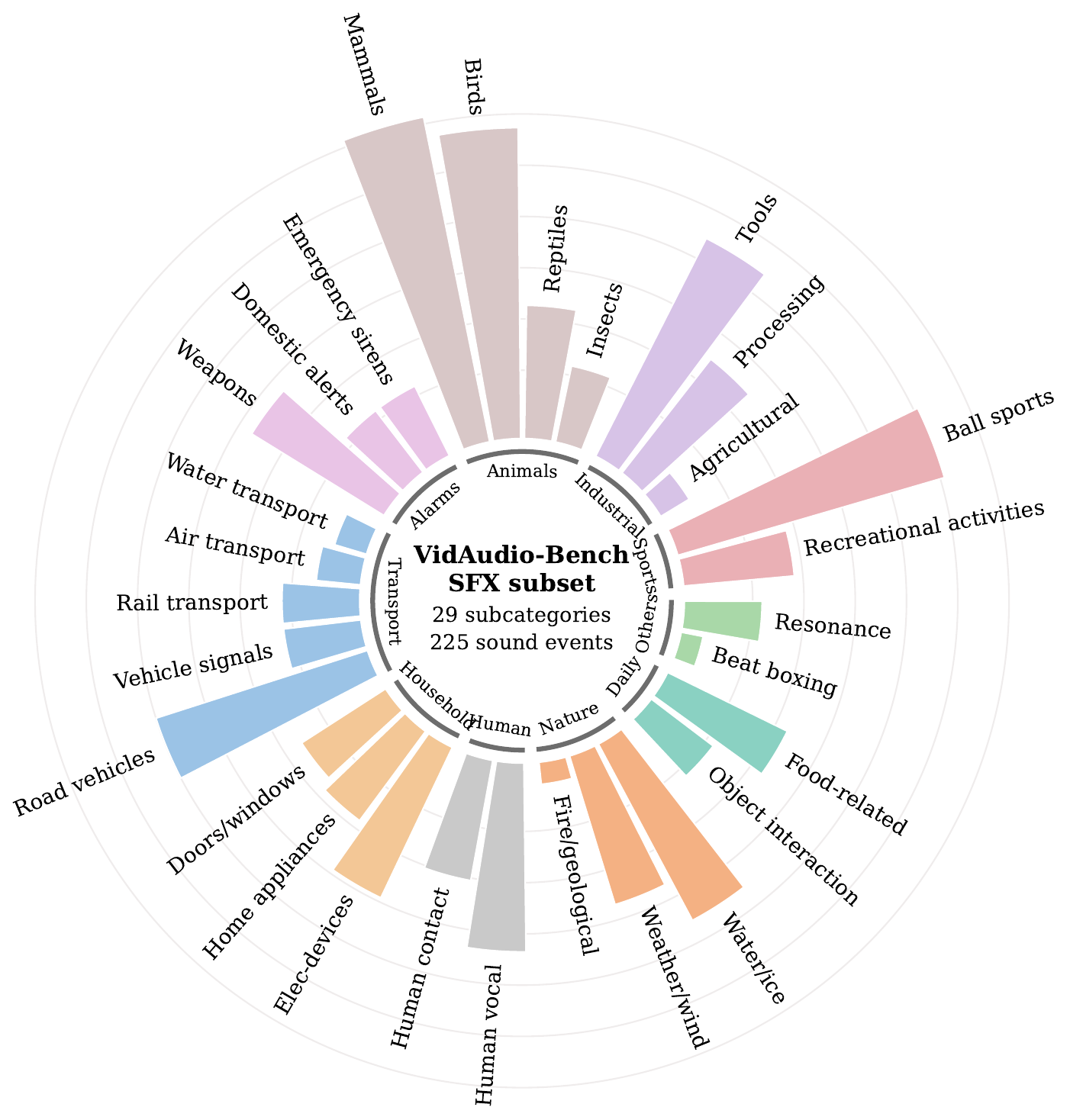}
  \caption{Data distribution of the SFX subset. The inner ring denotes high-level categories, and the outer bars show the number of sound events in each subcategory. }
    \label{SFX_subset}
\end{figure}

\begin{figure}[t]
  \centering

  \includegraphics[width=\linewidth]{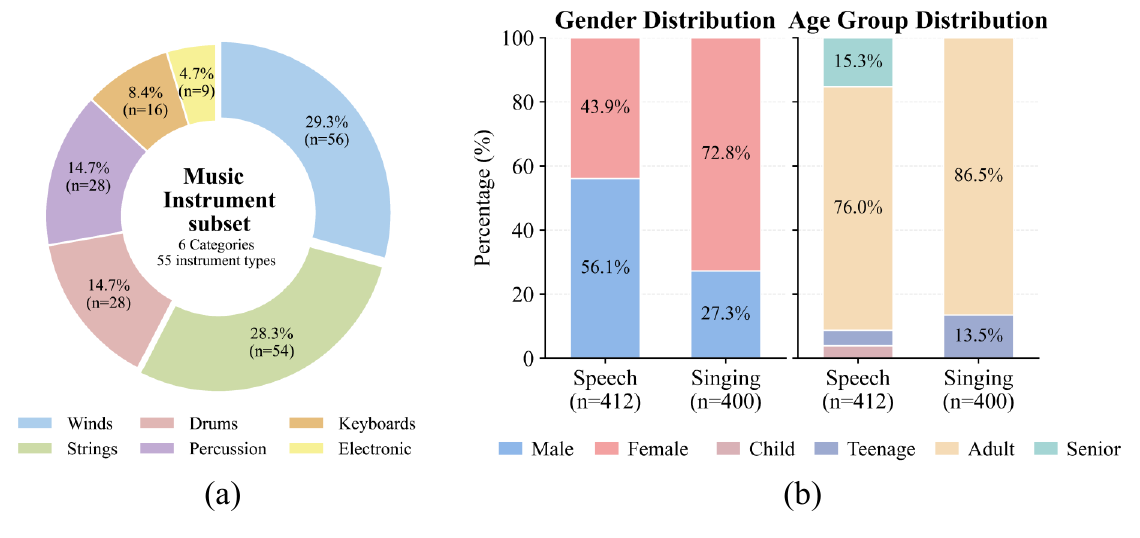}
  \caption{(a) Distribution of categories in the \textit{Instrument Performance }subset. (b) Gender and age group distributions in the \textit{Speech} and \textit{Singing }subsets. }
    \label{Music-idistribution}
\end{figure}

After this processing, the final benchmark comprises 1,634 high-quality video clips.  As shown in Figure~\ref{SFX_subset}, the \textit{SFX} subset encompasses 225 distinct sound events, spanning 10 major categories (Animals, Transport, Human Vocal, Sports, Household, Nature, Industrial, Alarms, Daily Activity, and Others) and 29 subcategories.
The \textit{Instrument Performance} subset includes 55 different instrument types, grouped into six categories: Strings, Winds, Percussion, Drums, Keyboards, and Electronic (Figure~\ref{Music-idistribution}(a)).
For the \textit{Speech} and \textit{Singing} subsets, we further analyze the apparent age and gender distributions of the visible subjects, as shown in Figure~\ref{Music-idistribution}(b). Further details are available in Appendix A.1.

\subsection{V2A and VT2A Paradigms}
\label{subsection:V2A and VT2A Paradigms}
A primary challenge in V2A generation is the one-to-many mapping between visual and acoustic signals. For instance, a beach scene could be paired with either ambient wave sounds or relaxing background music. This ambiguity complicates dimension-specific evaluations. To address this, we formalize two input configurations to ensure a controlled and category-specific assessment.

\noindent\textbf{V2A Setting (Task-Prompted):} 
To eliminate categorical ambiguity while strictly relying on visual cues for content generation, our baseline V2A setting employs a minimal task-level instruction (\textit{e.g.}, \textit{"Realistic foley sound synchronized with the video"}). This instruction serves merely as a categorical control signal, guiding the model toward the intended audio type without introducing explicit semantic descriptions of the visual events. We assess whether this minimal instruction is correctly followed through a dedicated \textit{Instruction-Following} dimension (see Section~\ref{subsection:Text-Audio Consistency Evaluation Metrics}).

\noindent\textbf{VT2A Setting (Caption-Augmented):} 
To investigate whether generation models truly understand the video content, we extend the benchmark to a Video-Text-to-Audio (VT2A) setting. In this setup, the input comprises the video and a prompt synthesized from visual content and task instructions. To prevent any potential information leakage from pre-existing audio-related text, we utilize Qwen3-VL~\cite{bai2025qwen3} to extract descriptions strictly from muted videos. This forces the Vision-Language Model (VLM) to describe only the visual elements (\textit{e.g.}, actions, objects, and environment). These raw visual descriptions are subsequently formatted to match specific audio generation templates, providing only video-observable semantics. This approach enables us to explore the model's generative ability when provided with visual text, and to assess whether this leads to meaningful improvements or provides semantic assistance, in comparison to the V2A setting.

We assess the fidelity of the generated visual descriptions using a hybrid validation strategy.
For tasks with original labels (\textit{SFX} and \textit{Instrument Performance}), LLM-based similarity evaluation yields a semantic retention accuracy of 70.4\% under a 0.5 threshold. For tasks without discrete labels (\textit{BGM}, \textit{Speech}, and \textit{Singing}), human evaluation on a 15\% sample shows strong alignment overall, with scores of 0.83 for \textit{Speech}, 0.81 for \textit{BGM}, and 0.67 for \textit{Singing}. These results confirm that our prompts provide effective visual semantics for VT2A generation. Detailed implementation of these two settings is provided in Appendix A.2.

\section{Evaluation Metrics}
\begin{figure*}[t]
  \centering
  \includegraphics[width=1\textwidth]{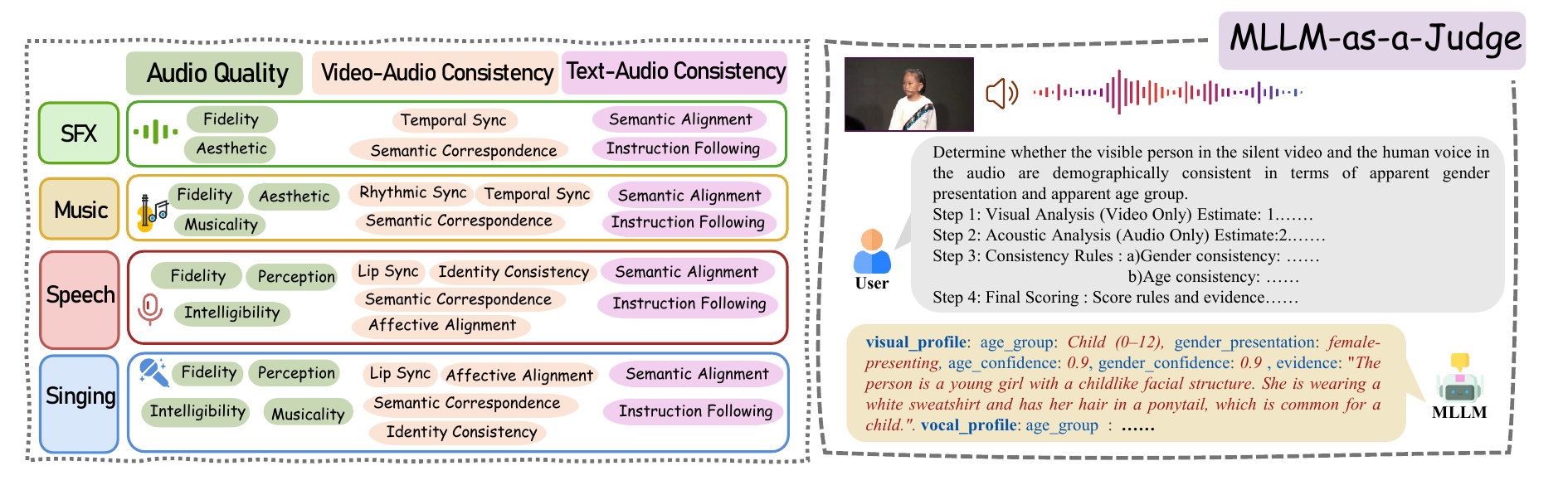}
  \caption{Overview of the evaluation framework in VidAudio-Bench. For each audio generation task, we define a set of evaluation dimensions across Audio quality, Video-Audio Consistency, and Text-Audio Consistency. An MLLM-as-a-Judge framework is employed to assess the generated audio through multi-step reasoning based on the input video and audio.}
  \label{fig:dimension_MLLM}
\end{figure*}

To comprehensively assess V2A generation, we propose a unified evaluation framework based on three complementary perspectives:
\textbf{(1) Audio Quality (AQ)}, which focuses on the intrinsic properties of the generated audio, including fidelity, perceptual quality, and task-specific characteristics (\textit{e.g.}, musicality). 
\textbf{(2) Video-Audio Consistency (VAC)}, which measures the alignment between audio and visual content in terms of semantic correspondence and temporal synchronization, as well as task-specific attributes such as affective alignment. 
\textbf{(3) Text-Audio Consistency (TAC)}, which assesses whether the generated audio conforms to the intended instruction or expected semantic content. These three perspectives are further refined into thirteen fine-grained dimensions, tailored to the characteristics of each task, as shown in Figure~\ref{fig:dimension_MLLM}.

\subsection{Audio Quality Evaluation}

\textbf{AQ - Fidelity.} To evaluate audio fidelity, signal integrity, and susceptibility to perceptual artifacts, we compute the Fréchet Distance (FD) on Audio-MAE~\cite{huang2022masked} embeddings. Compared to commonly used embeddings such as VGGish~\cite{roblek2019fr}, Audio-MAE demonstrates superior Precision sensitivity, providing an empirical upper bound for detecting additive noise and filtering artifacts~\cite{jeong2026empirical}. The reference distribution is constructed using category-matched real audio datasets. Specifically, for \textit{SFX} we use the sound event subset of VGG-Sound~\cite{chen2020vggsound}, while for \textit{Speech} we use the AVSpeech~\cite{ephrat2018looking} training set. This is necessary because FD measures distributional differences in the embedding space, and audio embeddings are highly dependent on semantic category and acoustic structure. Using mismatched reference distributions would cause FD to reflect content distribution differences rather than signal fidelity. Therefore, category-specific reference distributions allow FD to more accurately measure signal-level degradations while minimizing semantic distribution bias.

\noindent
\textbf{AQ - Aesthetic.}
We utilize the Audiobox-Aesthetics~\cite{tjandra2025meta} framework to assess the aesthetic quality of generated sound effects and music. This framework decomposes audio aesthetics into four key dimensions: Production Quality (PQ), Production Complexity (PC), Content Enjoyment (CE), and Content Usefulness (CU). For \textit{SFX} and \textit{Instrumental Performance} tasks, we prioritize PQ, CE, and CU as the primary metrics. PC is excluded in these cases because the visual context inherently constrains the audio’s degrees of freedom, rendering complexity a less meaningful indicator. In contrast, for \textit{BGM}, all four dimensions are considered. The weighting schemes—4:3:3 (CE:PQ:CU) for \textit{SFX/Instrument Performance} and 4:2:2:2 (CE:PQ:CU:PC) for \textit{BGM}—are based on utterance-level Pearson correlations in prior work~\cite{tjandra2025meta}, which we apply here for the first time to assign evaluation weights.

\noindent\textbf{AQ - Intelligibility.} 
Intelligibility is essential for evaluating \textit{Speech} and \textit{Singing}. In this work, we use STOI-Net~\cite{zezario2020stoi} to assess intelligibility in a non-intrusive manner. While STOI-Net has been previously applied to speech, we employ it here for the first time in singing generation. Traditional intrusive metrics such as STOI~\cite{taal2011algorithm}, which require access to the original clean waveform, are unsuitable for generative tasks~\cite{yeminilipvoicer}, as even a perfect model may not reproduce the original audio. Similarly, metrics like Character Error Rate (CER) and Word Error Rate (WER), which rely on reference transcriptions, are inapplicable for non-intrusive evaluation.

\noindent\textbf{AQ - Musicality.}
Following prior work~\cite{tian2025xmusic}, we evaluate musicality using three objective metrics: Pitch Class Histogram Entropy (PCE)~\cite{wu2020jazz} to quantify tonal clarity (where lower entropy indicates a more salient harmonic center), Grooving Pattern Similarity (GS)~\cite{wu2020jazz} to measure rhythmic regularity by comparing pattern consistency across bars, and Empty Beat Rate (EBR)~\cite{dong2018pypianoroll} to assess note density by calculating the proportion of silent beats. 
Since V2A models generate raw audio, we first convert the outputs to MIDI via Basic Pitch~\cite{bittner2022lightweight}. We define a Validity Rate ($V_{\mathrm{rate}}$) to account for samples yielding valid musical content.
We formulate \textbf{Musicality Score (MS)} in Eq.~\eqref{eq:final_score} by normalizing all metrics to $[0, 1]$, where $1 - \mathrm{PCE} / \log_2 12$ specifically quantifies tonal clarity relative to a uniform 12-pitch distribution.
\vspace{-1pt}
\begin{equation}
\mathrm{MS} = V_{\mathrm{rate}} \cdot \frac{\mathrm{GS} + \left(1 - \frac{\mathrm{PCE}}{\log_2 12}\right) + (1 - \mathrm{EBR})}{3}.
\label{eq:final_score}
\end{equation}

\noindent\textbf{AQ - Perception.}
To assess perceptual \textit{speech} quality in terms of naturalness, clarity, and overall listening experience, we employ DNSMOS Pro~\cite{cumlin2024dnsmos}, a non-intrusive probabilistic model for Mean Opinion Score (MOS) estimation. It adopts a lightweight end-to-end architecture to model the MOS posterior distribution, achieving high accuracy with reduced computational cost.
We use SingMOS-Pro~\cite{tang2025singmos} to evaluate the perceptual quality and acoustic pleasantness of the generated \textit{singing} voices. SingMOS-Pro is designed for automatic singing quality assessment, providing reliable MOS annotations of overall perceptual quality of singing vocals.

\subsection{Video-Audio Consistency Evaluation}

\textbf{VAC - Temporal Sync.}
We adopt the DeSync score predicted by the Synchformer \cite{iashin2024synchformer} to quantify event-level audio–video synchronization, where lower absolute values indicate better alignment. Following MMAudio \cite{cheng2025mmaudio}, we calculate the average offset using the first and last 4.8s of each video, allowing for overlap. This method evaluates both \textit{SFX} and \textit{Instrument Performance} tasks.

\noindent\textbf{VAC - Lip Sync.}
To evaluate audio-visual synchronization in \textit{Speech} and \textit{Singing}, we adopt LatentSync ~\cite{li2024latentsync}, which is designed for fine-grained lip-sync detection. We report the average absolute value of the offset, which measures temporal misalignment, where lower values indicate better synchronization.

\noindent\textbf{VAC - Rhythmic Sync.}
Standard tools like Synchformer are ill-suited for evaluating background music. Thus, we introduce a rhythmic synchronization score to jointly evaluate \textit{BGM }rhythm similarity and temporal alignment. The formula is
\vspace{-2pt} 
\begin{equation}
S_{\mathrm{rhythm}} = \frac{r+1}{2} \cdot \exp(-\alpha |\Delta t|),
\label{eq:rhythm}
\end{equation}
where \(r\) is the Pearson correlation coefficient between the video motion envelope and the audio energy envelope, and \(\Delta t\) is the optimal temporal offset estimated by cross-correlation. We map  \(r\) to \([0,1]\) before combining it with the temporal penalty term. The exponential factor penalizes large temporal offsets, where \(\alpha = \frac{\ln 2}{\tau}\), and \(\tau\) represents the time threshold at which the penalty weight halves.
Based on standard music theory and previous BGM generation practices~\cite{di2021video}, we \ set \(\tau = 0.5\) seconds, corresponding to one full beat at a tempo of 120 BPM. Such misalignment indicates the visual motion and musical rhythm are out of sync.
This formulation ensures that high synchronization scores require both strong correlation and minimal temporal misalignment, making it practical for reference-free evaluation of audio-visual rhythmic consistency.

\noindent\textbf{VAC - Semantic Correspondence.}
To evaluate the semantic consistency between visual and audio content, we employ $InternVL_{IB}^{\dagger}++(Ver.)$ from FreeBind~\cite{wang2024freebind}. By incorporating audio information from CLAP and fine-tuning the audio encoder, this space achieves improved audio–image alignment and strong performance across audio tasks. It provides a reliable metric for assessing audio–visual semantic consistency, surpassing the original ImageBind ~\cite{girdhar2023imagebind}.

\noindent\textbf{VAC - Identity Consistency.}
For \textit{Speech} and \textit{Singing} tasks, we adopt an MLLM-as-a-Judge framework (Figure~\ref{fig:dimension_MLLM}), using Qwen3-Omni~\cite{xu2025qwen3} as the judge model, to evaluate whether the generated voice is consistent with the visible person in the video. The evaluation focuses on demographic consistency, including apparent age and gender. Based on previous work~\cite{berthe2026s}, we categorize age groups into Child (0–12), Teenage (13–17), Adult (18–59), and Senior (60+). The detailed evaluation prompts are provided in Appendix B.1.

\noindent\textbf{VAC - Affective Alignment.}
Similarly, using the MLLM-as-a-Judge approach, we evaluate whether the emotion expressed in the generated \textit{speech} or \textit{singing} matches the emotion suggested by the visual scene. The detailed prompts are provided in Appendix B.2.

\subsection{Text-Audio Consistency Evaluation}
\label{subsection:Text-Audio Consistency Evaluation Metrics}
\noindent\textbf{TAC - Semantic Alignment.}
To evaluate the semantic consistency between textual content and generated audio, we adopt CLAP~\cite{wu2023large} to compute the cosine similarity of their embeddings. We use different checkpoints for different tasks to enhance the semantic alignment between audio and text representations.

\begin{table*}[!t]
\renewcommand{\arraystretch}{1} 
\centering
\caption{Evaluation results on VidAudio-Bench across multiple tasks and dimensions. Results are presented in the format of V2A / \textcolor{gray}{VT2A}, where $\uparrow$ ($\downarrow$) indicates higher (lower) values represent better performance. Best results in each category are highlighted in \textbf{bold}. \textit{Music-I}: Instrument Performance; \textit{Music-B}: Background Music.}
\label{comparison_task}
\footnotesize 

\begin{tabularx}{1\textwidth}{l | CCCCC CCC CC} 
\toprule
Dimensions and Tasks & \multicolumn{5}{c}{\textbf{Fidelity}$\downarrow$}& \multicolumn{3}{c}{\textbf{Aesthetic}$\uparrow$}  & \multicolumn{2}{c}{\textbf{Intelligibility}$\uparrow$} \\
\cmidrule(lr){2-6} \cmidrule(lr){7-9} \cmidrule(lr){10-11}
Models  & SFX  & Music-I &  Music-B & Speech & Singing & SFX  &  Music-I &  Music-B& Speech & Singing \\
\hline
AudioX~\cite{tian2025audiox} & \pair{5.480}{9.675}& \pair{10.533}{7.670} & \pair{25.297}{19.726} & \pair{7.300}{\textbf{5.607}}  & \pair{13.787}{\textbf{6.217}}& \pair{4.587}{4.785} & \pair{6.081}{6.133} & \pair{4.485}{5.503}& \pair{0.684}{0.626}& \pair{0.613}{0.593}  \\
FoleyCrafter~\cite{zhang2026foleycrafter}& \pair{11.847}{16.417}& \pair{17.537}{24.443} & \pair{20.100}{37.038} & \pair{19.362}{28.316}  & \pair{22.894}{27.529}& \pair{4.560}{4.951} & \pair{5.821}{5.892} & \pair{5.207}{4.369}& \pair{0.552}{0.556}& \pair{0.542}{0.528}  \\
HunyuanVideo-Foley~\cite{shan2025hunyuanvideo} & \pair{8.205}{12.240}& \pair{14.369}{7.721} & \pair{\textbf{10.187}}{22.904} & \pair{6.796}{21.030}  & \pair{58.529}{17.475}& \pair{\textbf{5.080}}{\textbf{5.060}} & \pair{6.189}{6.000} & \pair{\textbf{6.659}}{5.385}& \pair{0.583}{0.572}& \pair{0.557}{0.564}  \\
Kling-Foley~\cite{wang2025kling} & \pair{9.090}{9.529}& \pair{18.585}{17.114} & \pair{19.767}{\textbf{19.150}} & \pair{35.304}{35.005}  & \pair{12.981}{13.118}& \pair{4.855}{4.915} & \pair{5.957}{6.176} & \pair{6.035}{\textbf{6.006}}& \pair{0.554}{0.553}& \pair{0.552}{0.553} \\
MMAudio~\cite{cheng2025mmaudio} & \pair{13.499}{12.347}& \pair{17.952}{16.763} & \pair{24.060}{31.994} & \pair{14.402}{13.279}  & \pair{20.109}{16.004}& \pair{4.685}{4.740} & \pair{6.161}{6.002} & \pair{6.094}{4.937}& \pair{0.548}{0.534}& \pair{0.483}{0.468}  \\
ReWaS~\cite{jeong2025read}& \pair{10.308}{10.165}& \pair{17.429}{17.717} & \pair{34.675}{34.594} & \pair{23.848}{23.317}  & \pair{12.995}{13.007}& \pair{4.510}{4.525} & \pair{4.748}{4.728} & \pair{4.353}{4.348}& \pair{\textbf{0.861}}{\textbf{0.863}}& \pair{\textbf{0.896}}{\textbf{0.896}}  \\
ThinkSound~\cite{liuthinksound}& \pair{\textbf{4.943}}{\textbf{4.335}}& \pair{\textbf{6.582}}{\textbf{6.563}} & \pair{22.048}{20.562} & \pair{\textbf{4.490}}{7.343}  & \pair{\textbf{8.625}}{11.370}& \pair{4.653}{4.618} & \pair{\textbf{6.243}}{\textbf{6.282}} & \pair{6.047}{4.733}& \pair{0.599}{0.600}& \pair{0.572}{0.577}   \\
UniFlow-Audio~\cite{xu2025uniflow}& \pair{16.214}{17.484}& \pair{17.004}{27.911} & \pair{48.409}{54.746} & \pair{69.896}{68.009}  & \pair{51.483}{48.730}& \pair{4.745}{4.409} & \pair{5.534}{4.405} & \pair{3.982}{3.727}& \pair{0.528}{0.579}& \pair{0.424}{0.498}  \\

\bottomrule
\end{tabularx}

\vspace{0.1cm} 

\begin{tabularx}{1\textwidth}{l | CCCCC CC C CC} 
\toprule
Dimensions and Tasks & \multicolumn{5}{c}{\textbf{V-A Semantic-Corr}$\uparrow$}& \multicolumn{2}{c}{\textbf{Temp-Sync}$\downarrow$}  & \multicolumn{1}{c}{\textbf{Rhy-Sync}$\uparrow$} & \multicolumn{2}{c}{\textbf{Lip-Sync}$\downarrow$} \\
\cmidrule(lr){2-6} \cmidrule(lr){7-8} \cmidrule(lr){9-9} \cmidrule(lr){10-11}
Models  & SFX  &  Music-I &  Music-B & Speech & Singing & SFX  &  Music-I &  Music-B& Speech & Singing \\
\hline
AudioX~\cite{tian2025audiox} & \pair{0.194}{0.229}& \pair{0.238}{0.292} & \pair{0.120}{0.184} & \pair{0.157}{0.202}  & \pair{0.197}{0.215}& \pair{1.268}{1.246} & \pair{1.288}{1.343} & \pair{0.123}{0.103}& \pair{9.422}{9.809}& \pair{9.660}{9.115}  \\

FoleyCrafter~\cite{zhang2026foleycrafter}& \pair{0.201}{0.226}& \pair{0.205}{0.205} & \pair{0.179}{0.190} & \pair{0.164}{0.238}  & \pair{0.212}{0.241}& \pair{1.248}{1.227} & \pair{1.255}{1.297} & \pair{0.154}{0.132}& \pair{9.053}{8.916}& \pair{9.040}{9.184}  \\

HunyuanVideo-Foley~\cite{shan2025hunyuanvideo} & \pair{0.208}{0.224}& \pair{0.296}{0.294} & \pair{\textbf{0.187}}{\textbf{0.192}} & \pair{\textbf{0.238}}{\textbf{0.243}}  & \pair{0.228}{0.247}& \pair{0.673}{0.606} & \pair{0.375}{0.381} & \pair{0.134}{0.174}& \pair{2.351}{2.284}& \pair{1.024}{0.888}  \\

Kling-Foley~\cite{wang2025kling} & \pair{\textbf{0.234}}{\textbf{0.246}}& \pair{0.252}{0.283} & \pair{0.139}{0.149} & \pair{0.216}{0.218}  & \pair{\textbf{0.257}}{\textbf{0.259}}& \pair{0.526}{0.530} & \pair{0.394}{0.371} & \pair{0.139}{0.163}& \pair{2.062}{2.093}& \pair{0.853}{0.824} \\

MMAudio~\cite{cheng2025mmaudio} & \pair{0.191}{0.219}& \pair{\textbf{0.302}}{\textbf{0.316}} & \pair{0.159}{0.151} & \pair{0.220}{0.223}  & \pair{0.255}{0.232}& \pair{\textbf{0.480}}{\textbf{0.457}} & \pair{\textbf{0.260}}{\textbf{0.265}} & \pair{\textbf{0.172}}{0.187}& \pair{1.390}{1.464}& \pair{\textbf{0.733}}{\textbf{0.818}}  \\

ReWaS~\cite{jeong2025read}& \pair{0.072}{0.072}& \pair{0.059}{0.060} & \pair{0.076}{0.075} & \pair{0.047}{0.048}  & \pair{0.063}{0.063}& \pair{1.022}{1.036} & \pair{1.154}{1.158} & \pair{0.137}{0.138}& \pair{7.556}{7.809}& \pair{7.773}{7.746}  \\

ThinkSound~\cite{liuthinksound}& \pair{0.199}{0.199}& \pair{0.274}{0.289} & \pair{0.150}{0.156} & \pair{0.231}{0.232}  & \pair{0.232}{0.241}& \pair{0.620}{0.629} & \pair{0.382}{0.385} & \pair{0.163}{\textbf{0.192}}& \pair{\textbf{1.142}}{\textbf{1.151}}& \pair{0.968}{0.861}   \\

UniFlow-Audio~\cite{xu2025uniflow}& \pair{0.212}{0.166}& \pair{0.237}{0.171} & \pair{0.153}{0.121} & \pair{0.145}{0.122}  & \pair{0.139}{0.110}& \pair{1.127}{1.185} & \pair{1.267}{1.221} & \pair{0.148}{0.139}& \pair{9.600}{9.680}& \pair{9.131}{9.083}  \\

\bottomrule
\end{tabularx}

\vspace{0.1cm}

\begin{tabularx}{1\textwidth}{l | CCCCC CCCCC} 
\toprule
Dimensions and Tasks &\multicolumn{5}{c}{ \textbf{T-A Semantic-Align }$\uparrow$ } & \multicolumn{5}{c}{ \textbf{Instruction-Following}$\uparrow$}  \\
\cmidrule(lr){2-6} \cmidrule(lr){7-11}
Models & SFX  &  Music-I &  Music-B & Speech & Singing & SFX  &  Music-I &  Music-B & Speech &Singing  \\
\hline
AudioX~\cite{tian2025audiox} & \pair{0.032}{0.322}& \pair{0.231}{0.444} & \pair{0.235}{0.350} & \pair{0.341}{\textbf{0.437}}  & \pair{0.534}{0.414}& \pair{0.795}{0.832} & \pair{0.801}{0.843} & \pair{0.294}{0.537}& \pair{0.825}{0.981}& \pair{0.838}{\textbf{0.878}}  \\
FoleyCrafter~\cite{zhang2026foleycrafter}& \pair{0.028}{0.280}& \pair{0.288}{0.347} & \pair{0.293}{0.332} & \pair{0.323}{0.357}  & \pair{\textbf{0.549}}{0.365}& \pair{\textbf{0.973}}{0.965} & \pair{0.921}{0.812} & \pair{0.498}{0.195}& \pair{0.316}{\textbf{0.998}}& \pair{0.903}{0.580}  \\
HunyuanVideo-Foley~\cite{shan2025hunyuanvideo} & \pair{0.001}{0.329}& \pair{\textbf{0.343}}{0.447} & \pair{0.247}{\textbf{0.375}} & \pair{\textbf{0.404}}{0.329}  & \pair{0.455}{0.362}& \pair{0.475}{\textbf{0.988}} & \pair{\textbf{0.953}}{0.817} & \pair{0.723}{0.519}& \pair{0.993}{0.993}& \pair{0.930}{0.525}  \\
Kling-Foley~\cite{wang2025kling} & \pair{0.082}{0.320}& \pair{0.317}{\textbf{0.456}} & \pair{0.282}{0.281} & \pair{0.369}{0.400}  & \pair{0.478}{\textbf{0.416}}& \pair{0.935}{0.942} & \pair{0.753}{0.847} & \pair{0.530}{\textbf{0.704}}& \pair{0.959}{0.976}& \pair{0.828}{0.793} \\
MMAudio~\cite{cheng2025mmaudio} & \pair{\textbf{0.130}}{\textbf{0.352}}& \pair{0.263}{0.420} & \pair{\textbf{0.340}}{0.313} & \pair{0.269}{0.345}  & \pair{0.496}{0.338}& \pair{0.935}{0.978} & \pair{0.895}{0.837} & \pair{0.628}{0.355}& \pair{\textbf{0.995}}{\textbf{0.998}}& \pair{0.935}{0.553}  \\
ReWaS~\cite{jeong2025read}& \pair{0.001}{0.008}& \pair{0.274}{0.141} & \pair{0.173}{0.192} & \pair{0.278}{0.040}  & \pair{0.159}{0.152}& \pair{0.662}{0.658} & \pair{0.440}{0.450} & \pair{0.398}{0.390}& \pair{0.209}{0.209}& \pair{0.000}{0.003}  \\
ThinkSound~\cite{liuthinksound}& \pair{0.066}{0.213}& \pair{0.246}{0.382} & \pair{0.305}{0.215} & \pair{0.320}{0.361}  & \pair{0.478}{0.271}& \pair{0.848}{0.818} & \pair{0.911}{\textbf{0.916}} & \pair{\textbf{0.736}}{0.316}& \pair{0.956}{0.932}& \pair{\textbf{0.960}}{0.835}   \\
UniFlow-Audio~\cite{xu2025uniflow}& \pair{0.019}{0.152}& \pair{0.335}{0.315} & \pair{0.119}{0.195} & \pair{0.348}{0.285}  & \pair{0.177}{0.263}& \pair{0.920}{0.928} & \pair{0.880}{0.723} & \pair{0.052}{0.039}& \pair{0.711}{0.515}& \pair{0.050}{0.040}  \\

\bottomrule
\end{tabularx}

\vspace{0.1cm}

\begin{tabularx}{1\textwidth}{l | CCCCC CCCCC} 
\toprule
Dimensions and Tasks& \multicolumn{3}{c}{\textbf{Musicality}$\uparrow$} & \multicolumn{2}{c}{\textbf{Perception}$\uparrow$}& \multicolumn{2}{c}{\textbf{Identity-Cons}$\uparrow$} & \multicolumn{2}{c}{\textbf{Affective-Align}$\uparrow$}  \\
\cmidrule(lr){2-4}
\cmidrule(lr){5-6}
\cmidrule(lr){7-8}
\cmidrule(lr){9-10}
Models & Music-I & Music-B & Singing&  Speech & Singing & Speech & Singing  & Speech & Singing  & Avg.Rank\\
\hline
AudioX~\cite{tian2025audiox} & \pair{0.637}{0.714}& \pair{0.700}{0.716} & \pair{0.723}{\textbf{0.741}} & \pair{2.072}{2.217}  & \pair{2.762}{2.853}& \pair{4.388}{4.663} & \pair{3.405}{3.825}  & \pair{2.187}{2.345}& \pair{3.233}{3.203} & \pair{5.462}{3.769} \\
FoleyCrafter~\cite{zhang2026foleycrafter}& \pair{0.713}{0.635}& \pair{0.646}{0.565} & \pair{\textbf{0.763}}{0.674}& \pair{2.271}{2.869}& \pair{2.091}{2.365} & \pair{3.325}{4.607} & \pair{3.943}{3.138} & \pair{2.403}{3.034}& \pair{3.448}{3.343} & \pair{4.795}{5.179} \\
HunyuanVideo-Foley~\cite{shan2025hunyuanvideo} & \pair{\textbf{0.751}}{0.652}& \pair{\textbf{0.751}}{0.668} & \pair{0.735}{0.683}& \pair{\textbf{3.037}}{\textbf{2.924}}& \pair{3.212}{3.307} & \pair{\textbf{4.901}}{4.908} & \pair{3.798}{4.325} & \pair{2.420}{2.437}& \pair{3.748}{3.795}  & \pair{3.128}{3.359}\\
Kling-Foley~\cite{wang2025kling}  & \pair{0.614}{0.669}& \pair{0.715}{\textbf{0.730}} & \pair{0.694}{0.705}& \pair{2.588}{2.591}& \pair{3.513}{\textbf{3.507}} & \pair{4.818}{4.823} & \pair{3.928}{3.925} & \pair{2.762}{\textbf{2.697}}& \pair{3.943}{3.885} & \pair{3.718}{3.128} \\
MMAudio~\cite{cheng2025mmaudio}  & \pair{0.678}{0.656}& \pair{0.703}{0.616} & \pair{0.748}{0.642} & \pair{2.983}{2.874}& \pair{\textbf{3.625}}{3.365} & \pair{4.784}{\textbf{4.944}} & \pair{4.128}{\textbf{4.545}} & \pair{2.204}{2.488}& \pair{3.808}{3.690} & \pair{3.436}{3.769} \\
ReWaS~\cite{jeong2025read}& \pair{0.545}{0.551}& \pair{0.579}{0.578} & \pair{0.605}{0.606}& \pair{2.048}{2.046}& \pair{1.458}{1.460} & \pair{3.782}{3.842} & \pair{3.348}{3.265} & \pair{2.354}{2.318}& \pair{2.920}{2.943}  & \pair{6.513}{6.590}\\
ThinkSound~\cite{liuthinksound}& \pair{0.716}{\textbf{0.726}}& \pair{0.748}{0.678} & \pair{0.746}{0.710} & \pair{2.689}{2.599}& \pair{3.005}{3.039} & \pair{4.626}{4.614} & \pair{3.170}{3.475} & \pair{\textbf{2.833}}{2.648}& \pair{3.873}{\textbf{4.085}}   & \pair{3.051}{3.513}\\
UniFlow-Audio~\cite{xu2025uniflow}& \pair{0.704}{0.694}& \pair{0.542}{0.515} & \pair{0.673}{0.651} & \pair{1.721}{1.679}& \pair{1.993}{2.005} & \pair{4.473}{4.478} & \pair{\textbf{4.205}}{4.263} & \pair{2.726}{2.408}& \pair{\textbf{4.175}}{3.975}  & \pair{5.820}{6.641}\\

\bottomrule
\end{tabularx}
\end{table*}

\noindent\textbf{TAC - Instruction Following.}
As defined in Section~\ref{subsection:V2A and VT2A Paradigms}, each generation task is guided by a specific category instruction. It is crucial to assess whether the generation model faithfully follows the given instruction and produces audio of the intended category. To this end, we adopt the MLLM-as-a-Judge framework to systematically verify instruction compliance. More detailed implementation information of the judge model can be found in Appendix B.3.

\section{Experiments}
\subsection{Evaluated Audio Generation Models}
We evaluate 11 representative models on VidAudio-Bench, including 8 video-to-audio models and 3 video-to-music (V2M) models, covering 10 open-source models and 1 commercial model. The evaluated models are briefly summarized as follows:
\begin{itemize}[leftmargin=*,nosep]
\item \textbf{AudioX}~\cite{tian2025audiox} is a unified Diffusion Transformer (DiT) for anything-to-audio generation that supports diverse multimodal conditions, including video, text, and images.
\item \textbf{FoleyCrafter}~\cite{zhang2026foleycrafter} adapts a pre-trained T2A model for V2A generation with a semantic adapter and a temporal controller.
\item \textbf{HunyuanVideo-Foley}~\cite{shan2025hunyuanvideo} is an end-to-end VT2A DiT that leverages self-supervised audio features and dual-stream fusion for high-fidelity synchronization.
\item \textbf{Kling-Foley}~\cite{wang2025kling} is a DiT-based V2A model enhancing visual-semantic and temporal alignment for high-fidelity synthesis.
\item \textbf{MMAudio}~\cite{cheng2025mmaudio} is a multimodal framework improving V2A synthesis by jointly learning from text-audio and audio-visual data.
\item \textbf{ReWaS}~\cite{jeong2025read} is a VT2A method that uses video as structural control and text prompts as semantic guidance.
\item \textbf{ThinkSound}~\cite{liuthinksound} introduces Chain-of-Thought reasoning into V2A generation for stepwise audio synthesis and editing.
\item \textbf{UniFlow-Audio}~\cite{xu2025uniflow} is a unified flow-matching framework employing a dual-fusion mechanism for omni-modal alignment.
\item \textbf{GVMGen}~\cite{zuo2025gvmgen} is a V2M model using hierarchical attention for spatial-temporal alignment in zero-shot music generation.
\item \textbf{SONIQUE}~\cite{zhang2025sonique} is a customizable V2M model that uses LLMs to bridge unpaired data by converting visual descriptions into musical tags for diffusion-based generation.
\item \textbf{VidMuse}~\cite{tian2025vidmuse} is a V2M framework with long-short-term modeling capturing both local and global visual cues.
\end{itemize}

\begin{figure}[b]
  \centering
  \includegraphics[width=0.9\linewidth]{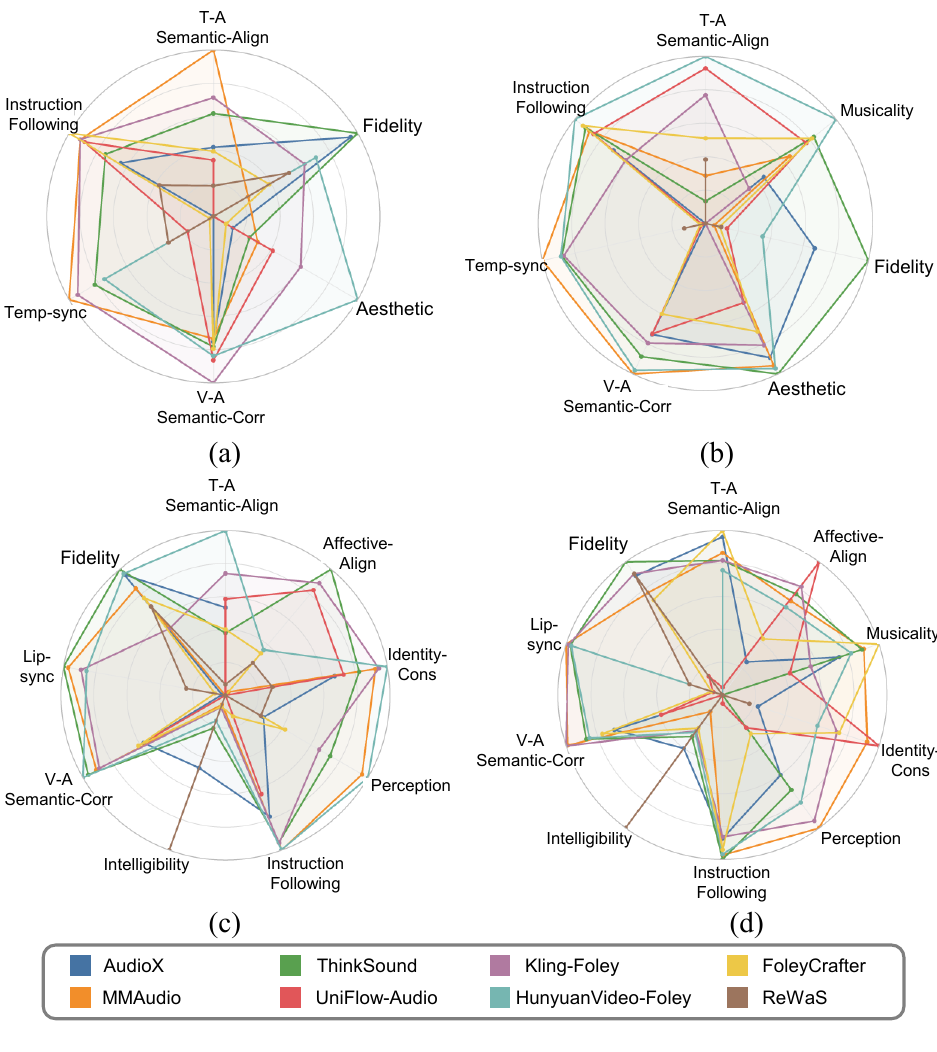}
  \caption{Task-wise radar plots of eight representative models across four audio generation tasks using V2A results: (a) \textit{sound effects}, (b) \textit{music}, (c) \textit{speech}, and (d)\textit{ singing}. Each subplot summarizes model performance over the task-specific evaluation dimensions.} 
  \label{radio_pic}
\end{figure}

\subsection{Main Results}
Table~\ref{comparison_task} presents the performance of all models on VidAudio-Bench.

\noindent\textbf{Overall Findings}. Model performance varies substantially across tasks and evaluation dimensions. Although many models perform reasonably well on \textit{SFX} and \textit{Music}, \textit{Speech} and \textit{Singing} remain much more challenging, with clear drops in perceptual quality. One likely reason is that current mainstream V2A training data (\textit{e.g.}, VGG-Sound~\cite{chen2020vggsound} and AudioSet~\cite{gemmeke2017audio}) are dominated by environmental sound events, providing limited coverage of highly structured and semantically complex human vocalizations.

We also observe that no single model consistently ranks best across all perspectives and dimensions. Instead, different models exhibit different strengths, suggesting inherent trade-offs among these objectives. For example, models that align better with visual content (\textit{e.g.}, Kling-Foley) do not necessarily generate higher-quality audio, while models with stronger perceptual quality (\textit{e.g.}, AudioX) often fall short on fine-grained temporal alignment, such as lip synchronization. These results suggest that V2A/VT2A evaluation cannot be adequately captured by a single overall score, and instead requires a multi-dimensional, task-aware evaluation framework.

\noindent\textbf{Task-wise Results}. We conduct detailed analyses across more than four task categories, as illustrated in Figure~\ref{radio_pic}.

\textit{Sound Effects}. As shown in Figure~\ref{radio_pic}(a), ThinkSound and AudioX lead in fidelity, MMAudio excels in temporal synchronization and text–audio consistency, while Kling achieves stronger video–audio alignment. All the models demonstrate relatively good performance.

\textit{Music}. Figure~\ref{radio_pic}(b) shows that for the \textit{Instrument Performance} task, ThinkSound leads in fidelity and aesthetic quality, with melody performance second only to Hunyuan. MMAudio excels in temporal synchronization and video–audio alignment, while Hunyuan achieves the best text–audio consistency. Most models also successfully follow the instructions to generate the intended music.
For the \textit{BGM} task, since most general-purpose models struggle with background music generation, we additionally compare three specialized BGM generation models (Table~\ref{BGM_comparison}). Hunyuan achieves the highest fidelity, followed by GVMGen and VidMuse, and exhibits stronger melodic structure than GVMGen. In terms of rhythmic synchronization, Kling performs the best. However, these specialized models score lower on semantic consistency, indicating that optimizing for musical quality alone does not guarantee strong semantic alignment with visual or textual conditions.
\begin{table}
\centering
\caption{Evaluation results of three specialized BGM generation models on the \textit{BGM} task across five dimensions.}
\label{BGM_comparison}
\renewcommand\arraystretch{1.15} 
\resizebox{0.48\textwidth}{!}{
\setlength{\tabcolsep}{4pt} 

\begin{tabular}{l ccccc} 
\toprule
Models & Fidelity$\downarrow$ & Aesthetic$\uparrow$ & Musicality$\uparrow$ & Semantic-Corr$\uparrow$ & Rhy-Sync$\uparrow$ \\
\midrule 
GVMGen~\cite{zuo2025gvmgen} & 12.171 & 6.598 & \textbf{0.746} & 0.100 & \textbf{0.151} \\
SONIQUE~\cite{zhang2025sonique} & 34.807 & 5.767 & 0.638 & 0.113 & 0.093 \\
VidMuse~\cite{tian2025vidmuse} & \textbf{10.960} & \textbf{7.080 }& 0.717 & \textbf{0.141} & 0.128 \\
\bottomrule
\end{tabular}}
\end{table}

\begin{figure*}[t]
  \centering
  \includegraphics[width=0.95\textwidth]{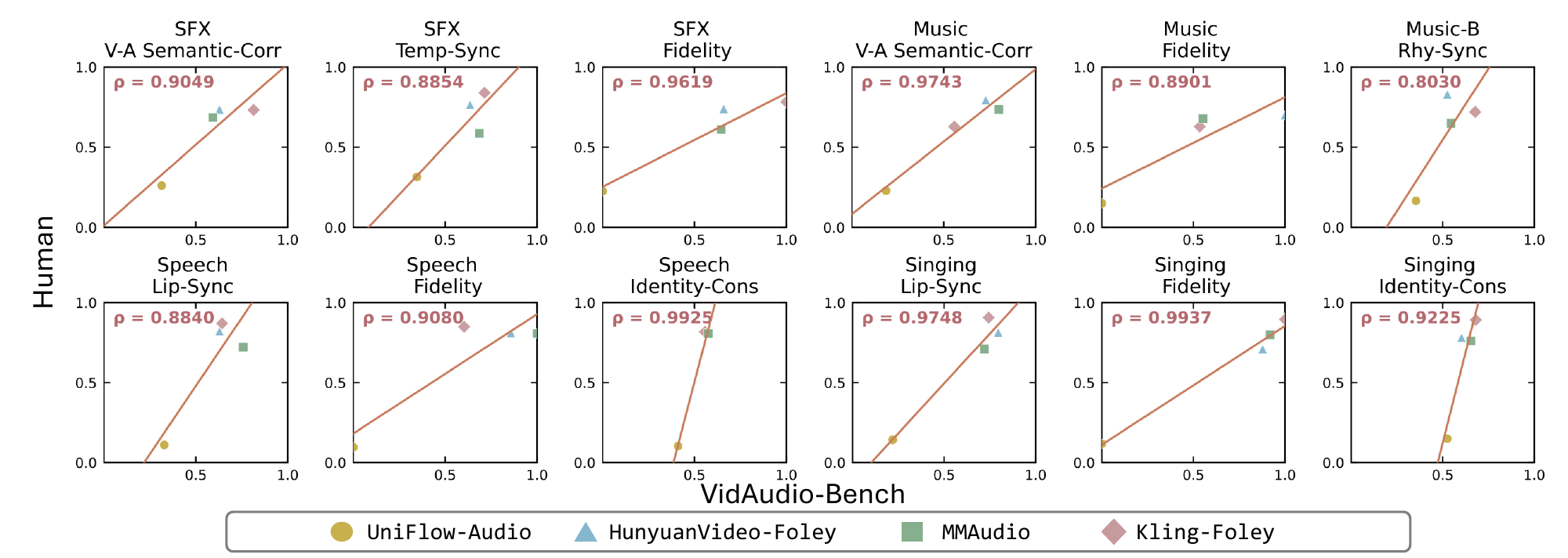}
  \caption{Human preference correlation with VidAudio-Bench. This figure shows the Pearson correlation coefficients ($\rho$)  between VidAudio-Bench scores (x-axis) and human win rates (y-axis) across various evaluation dimensions. High correlations demonstrate strong alignment with human perceptual judgments.
  }
  \label{fig:pearson}
\end{figure*}

\textit{Speech}. As shown in Figure~\ref{radio_pic}(c), several models (\textit{e.g.}, ThinkSound) achieve strong performance in lip synchronization, fidelity, and feature consistency. Although ReWaS achieves the highest intelligibility, its remarkably low Instruction-Following score indicates a critical qualitative flaw. Specifically, while the model produces perfectly articulated syllables, they form nonsensical gibberish when strung together, leading the MLLM to reject the audio as natural human speech. These results reveal a clear trilemma in current speech generation models: jointly optimizing clarity, synchronization, and semantic alignment remains challenging.

\textit{Singing}. As shown in Figure~\ref{radio_pic}(d), MMAudio achieves strong overall vocal perceptual quality, with melody performance second only to FoleyCrafter, and also excels in synchronization but lags behind most models in intelligibility. Kling demonstrates stronger overall semantic consistency. Notably, no existing model can simultaneously balance melodicity, lyric intelligibility, visual synchronization, and semantic alignment in singing tasks.

\noindent\textbf{V2A vs. VT2A}. To investigate the effect of explicit visual descriptions on generation, we jointly analyze video-audio semantic correspondence (V-A Semantic-Corr) and instruction following (IF), where IF measures whether the generated audio matches the target category. As shown in Table~\ref{comparison_task}, rather than improving, IF frequently drops when moving from the V2A setting to VT2A. This drop is particularly evident in complex audio categories such as \textit{BGM} and \textit{Singing}. For instance, MMAudio drops from 0.935 to 0.553 on \textit{Singing}, and HunyuanVideo-Foley drops from 0.930 to 0.525.

We attribute this degradation to two factors. First, longer visual descriptions can dilute the core instruction, making it harder for the model to preserve the target category when processing dense contextual details. As a result, the model may be distracted by secondary information, such as objects, attributes, or scene elements, instead of following the main task requirement. Second, explicit visual descriptions may introduce semantic cues that conflict with the intended audio category. For instance, in the \textit{BGM} task, descriptions of visible actions such as \textit{"a car speeding"} can bias the model toward generating event-driven sound effects rather than background music, leading to severe IF drops, such as UniFlow-Audio falling to 0.039 on \textit{BGM}. Overall, these results point to a fundamental tension in current V2A systems between category-level control and visually grounded generation. This also explains why some models achieve higher V-A semantic consistency in VT2A despite lower IF: even when they miss the intended category, the generated audio can still align closely with the visible content of the video.

\subsection{Human Evaluation Correlation Analysis}
In this section, we conduct large-scale human evaluations and compute the correlation between scores from automatic metrics and humans to validate VidAudio-Bench’s alignment with human senses.

\noindent\textbf{Human Evaluation.} For each task, we selected 20 representative videos. We then paired these with audio generated by 4 different models, resulting in a total of 400 audio–video pairs. To mitigate the influence of individual subjective preferences, each pair was evaluated by five independent raters. To avoid cross-task interference, each rater was assigned to evaluate only one specific type of audio task. In total, 20 participants were involved in the study. For each audio category, we focus on four key dimensions: semantics, synchronization, realism, and instruction following. After watching each video, raters scored from 1 to 5 for each of these dimensions.

\noindent\textbf{Evaluation Methodology.} We employ three strategies to assess alignment with human preferences: \textit{(1) Pairwise Correlation:} For metrics allowing pairwise comparison, we calculate win rates, assigning 1 for a win, 0 for a loss, and 0.5 for a tie, for both human and benchmark scores, then compute the Pearson correlation ($\rho$) between them. (\textit{2) Direct Correlation:} For metrics like Fidelity, we directly correlate raw model scores with human ratings. As shown in Figure~\ref{fig:pearson}, these high correlations demonstrate strong alignment with human judgments. \textit{(3) Classification Accuracy:} For the IF metric, we adopted a binary classification evaluation to assess the agreement between human-labeled categories and the MLLM's predictions. The results are summarized in Table~\ref{IF_acc}. High accuracy and F1-scores across all categories confirm that our automated IF metric closely mirrors human judgment in verifying instruction adherence.

\begin{table}
\centering
\caption{Binary classification performance of the automated \textit{Instruction Following} metric against human-labeled audio categories across four task types.}
\label{IF_acc}
\renewcommand\arraystretch{1.0} 
\resizebox{0.38\textwidth}{!}{
\setlength{\tabcolsep}{4pt} 

\begin{tabular}{l|cccc}
\toprule
Category & Accuracy$\uparrow$ & Precision$\uparrow$ & Recall$\uparrow$ & F1-score$\uparrow$ \\
\midrule 
SFX     & 0.8625 & 0.8591 & 0.9922 & 0.9209 \\
Music   & 0.8000 & 0.7253 & 0.9041 & 0.8042 \\
Speech  & 0.8875 & 0.8855 & 0.9748 & 0.9280 \\
Singing & 0.8438 & 0.8000 & 0.9268 & 0.8588 \\
\bottomrule
\end{tabular}}
\end{table}

\section{Conclusion}

In conclusion, VidAudio-Bench establishes a comprehensive benchmark for V2A/VT2A evaluation, built on 1,634 carefully curated video-text pairs with strong audio-visual correlation. Covering four audio types and thirteen evaluation dimensions, it supports reliable and interpretable assessment through automated, multidimensional, and human-aligned evaluation. The benchmark also reveals the central challenge of balancing Audio Quality, Video-Audio Consistency, and Text-Audio Consistency. VidAudio-Bench offers valuable insights for achieving more coherent and perceptually grounded audio generation, constituting a significant and robust contribution to research and evaluation in this field.

\bibliographystyle{ACM-Reference-Format}
\bibliography{sample-base}

@String{Computer = "{IEEE} Computer" }

@String{Springer = "Springer-Verlag" }

@inproceedings{roblek2019fr,
  title={Fr$\backslash$'echet Audio Distance: A Reference-free Metric for Evaluating Music Enhancement Algorithms},
  author={Roblek, Dominik and Kilgour, Kevin and Sharifi, Matt and Zuluaga, Mauricio},
  booktitle={Proc. Interspeech},
  pages={2350--2354},
  year={2019}
}

@article{tjandra2025meta,
  title={Meta audiobox aesthetics: Unified automatic quality assessment for speech, music, and sound},
  author={Tjandra, Andros and Wu, Yi-Chiao and Guo, Baishan and Hoffman, John and Ellis, Brian and Vyas, Apoorv and Shi, Bowen and Chen, Sanyuan and Le, Matt and Zacharov, Nick and others},
  journal={arXiv preprint arXiv:2502.05139},
  year={2025}
}

@article{jeong2026empirical,
  title={An Empirical Analysis of Task-Induced Encoder Bias in Fr$\backslash$'echet Audio Distance},
  author={Jeong, Wonwoo},
  journal={arXiv preprint arXiv:2602.23958},
  year={2026}
}

@article{huang2022masked,
  title={Masked autoencoders that listen},
  author={Huang, Po-Yao and Xu, Hu and Li, Juncheng and Baevski, Alexei and Auli, Michael and Galuba, Wojciech and Metze, Florian and Feichtenhofer, Christoph},
  journal={Advances in neural information processing systems},
  volume={35},
  pages={28708--28720},
  year={2022}
}

@inproceedings{zezario2020stoi,
  title={STOI-Net: A deep learning based non-intrusive speech intelligibility assessment model},
  author={Zezario, Ryandhimas E and Fu, Szu-Wei and Fuh, Chiou-Shann and Tsao, Yu and Wang, Hsin-Min},
  booktitle={2020 Asia-Pacific Signal and Information Processing Association Annual Summit and Conference (APSIPA ASC)},
  pages={482--486},
  year={2020},
  organization={IEEE}
}

@inproceedings{yeminilipvoicer,
  title={LipVoicer: Generating Speech from Silent Videos Guided by Lip Reading},
  author={Yemini, Yochai and Shamsian, Aviv and Bracha, Lior and Gannot, Sharon and Fetaya, Ethan},
  booktitle={The Twelfth International Conference on Learning Representations}
}

@inproceedings{cumlin2024dnsmos,
  title={DNSMOS Pro: A Reduced-Size DNN for Probabilistic MOS of Speech},
  author={Cumlin, Fredrik and Liang, Xinyu and Ungureanu, Victor and Reddy, Chandan KA and Sch{\"u}ldt, Christian and Chatterjee, Saikat},
  booktitle={Interspeech},
  year={2024}
}

@article{tian2025xmusic,
  title={Xmusic: Towards a generalized and controllable symbolic music generation framework},
  author={Tian, Sida and Zhang, Can and Yuan, Wei and Tan, Wei and Zhu, Wenjie},
  journal={IEEE Transactions on Multimedia},
  volume={27},
  pages={6857--6871},
  year={2025},
  publisher={IEEE}
}

@article{wu2020jazz,
  title={The jazz transformer on the front line: Exploring the shortcomings of ai-composed music through quantitative measures},
  author={Wu, Shih-Lun and Yang, Yi-Hsuan},
  journal={arXiv preprint arXiv:2008.01307},
  year={2020}
}

@article{dong2018pypianoroll,
  title={Pypianoroll: Open source Python package for handling multitrack pianoroll},
  author={Dong, Hao-Wen and Hsiao, Wen-Yi and Yang, Yi-Hsuan},
  journal={Proc. ISMIR. Late-breaking paper},
  year={2018}
}

@article{tang2025singmos,
  title={SingMOS-Pro: An Comprehensive Benchmark for Singing Quality Assessment},
  author={Tang, Yuxun and Liu, Lan and Feng, Wenhao and Zhao, Yiwen and Han, Jionghao and Yu, Yifeng and Shi, Jiatong and Jin, Qin},
  journal={arXiv preprint arXiv:2510.01812},
  year={2025}
}

@inproceedings{bittner2022lightweight,
  title={A lightweight instrument-agnostic model for polyphonic note transcription and multipitch estimation},
  author={Bittner, Rachel M and Bosch, Juan Jos{\'e} and Rubinstein, David and Meseguer-Brocal, Gabriel and Ewert, Sebastian},
  booktitle={ICASSP 2022-2022 IEEE International Conference on Acoustics, Speech and Signal Processing (ICASSP)},
  pages={781--785},
  year={2022},
  organization={IEEE}
}

@inproceedings{iashin2024synchformer,
  title={Synchformer: Efficient synchronization from sparse cues},
  author={Iashin, Vladimir and Xie, Weidi and Rahtu, Esa and Zisserman, Andrew},
  booktitle={ICASSP 2024-2024 IEEE International Conference on Acoustics, Speech and Signal Processing (ICASSP)},
  pages={5325--5329},
  year={2024},
  organization={IEEE}
}

@inproceedings{cheng2025mmaudio,
  title={Mmaudio: Taming multimodal joint training for high-quality video-to-audio synthesis},
  author={Cheng, Ho Kei and Ishii, Masato and Hayakawa, Akio and Shibuya, Takashi and Schwing, Alexander and Mitsufuji, Yuki},
  booktitle={Proceedings of the Computer Vision and Pattern Recognition Conference},
  pages={28901--28911},
  year={2025}
}

@article{li2024latentsync,
  title={Latentsync: Taming audio-conditioned latent diffusion models for lip sync with syncnet supervision},
  author={Li, Chunyu and Zhang, Chao and Xu, Weikai and Lin, Jingyu and Xie, Jinghui and Feng, Weiguo and Peng, Bingyue and Chen, Cunjian and Xing, Weiwei},
  journal={arXiv preprint arXiv:2412.09262},
  year={2024}
}

@inproceedings{wang2024freebind,
  title={FreeBind: free lunch in unified multimodal space via knowledge fusion},
  author={Wang, Zehan and Zhang, Ziang and Cheng, Xize and Huang, Rongjie and Liu, Luping and Ye, Zhenhui and Huang, Haifeng and Zhao, Yang and Jin, Tao and Gao, Peng and others},
  booktitle={Proceedings of the 41st International Conference on Machine Learning},
  pages={52233--52246},
  year={2024}
}

@inproceedings{girdhar2023imagebind,
  title={Imagebind: One embedding space to bind them all},
  author={Girdhar, Rohit and El-Nouby, Alaaeldin and Liu, Zhuang and Singh, Mannat and Alwala, Kalyan Vasudev and Joulin, Armand and Misra, Ishan},
  booktitle={Proceedings of the IEEE/CVF conference on computer vision and pattern recognition},
  pages={15180--15190},
  year={2023}
}

@article{xu2025qwen3,
  title={Qwen3-omni technical report},
  author={Xu, Jin and Guo, Zhifang and Hu, Hangrui and Chu, Yunfei and Wang, Xiong and He, Jinzheng and Wang, Yuxuan and Shi, Xian and He, Ting and Zhu, Xinfa and others},
  journal={arXiv preprint arXiv:2509.17765},
  year={2025}
}

@inproceedings{wu2023large,
  title={Large-scale contrastive language-audio pretraining with feature fusion and keyword-to-caption augmentation},
  author={Wu, Yusong and Chen, Ke and Zhang, Tianyu and Hui, Yuchen and Berg-Kirkpatrick, Taylor and Dubnov, Shlomo},
  booktitle={ICASSP 2023-2023 IEEE International Conference on Acoustics, Speech and Signal Processing (ICASSP)},
  pages={1--5},
  year={2023},
  organization={IEEE}
}

@inproceedings{kreukaudiogen,
  title={AudioGen: Textually Guided Audio Generation},
  author={Kreuk, Felix and Synnaeve, Gabriel and Polyak, Adam and Singer, Uriel and D{\'e}fossez, Alexandre and Copet, Jade and Parikh, Devi and Taigman, Yaniv and Adi, Yossi},
  booktitle={The Eleventh International Conference on Learning Representations}
}

@article{yang2023diffsound,
  title={Diffsound: Discrete diffusion model for text-to-sound generation},
  author={Yang, Dongchao and Yu, Jianwei and Wang, Helin and Wang, Wen and Weng, Chao and Zou, Yuexian and Yu, Dong},
  journal={IEEE/ACM Transactions on Audio, Speech, and Language Processing},
  volume={31},
  pages={1720--1733},
  year={2023},
  publisher={IEEE}
}

@inproceedings{yang2021multi,
  title={Multi-band melgan: Faster waveform generation for high-quality text-to-speech},
  author={Yang, Geng and Yang, Shan and Liu, Kai and Fang, Peng and Chen, Wei and Xie, Lei},
  booktitle={2021 IEEE Spoken Language Technology Workshop (SLT)},
  pages={492--498},
  year={2021},
  organization={IEEE}
}

@inproceedings{huang2023make,
  title={Make-an-audio: Text-to-audio generation with prompt-enhanced diffusion models},
  author={Huang, Rongjie and Huang, Jiawei and Yang, Dongchao and Ren, Yi and Liu, Luping and Li, Mingze and Ye, Zhenhui and Liu, Jinglin and Yin, Xiang and Zhao, Zhou},
  booktitle={International Conference on Machine Learning},
  pages={13916--13932},
  year={2023},
  organization={PMLR}
}

@inproceedings{melechovsky2024mustango,
  title={Mustango: Toward controllable text-to-music generation},
  author={Melechovsky, Jan and Guo, Zixun and Ghosal, Deepanway and Majumder, Navonil and Herremans, Dorien and Poria, Soujanya},
  booktitle={Proceedings of the 2024 Conference of the North American Chapter of the Association for Computational Linguistics: Human Language Technologies (Volume 1: Long Papers)},
  pages={8293--8316},
  year={2024}
}

@inproceedings{zivmasked,
  title={Masked Audio Generation using a Single Non-Autoregressive Transformer},
  author={Ziv, Alon and Gat, Itai and Le Lan, Gael and Remez, Tal and Kreuk, Felix and Copet, Jade and D{\'e}fossez, Alexandre and Synnaeve, Gabriel and Adi, Yossi},
  booktitle={The Twelfth International Conference on Learning Representations}
}

@article{chen2024images,
  title={Images that sound: Composing images and sounds on a single canvas},
  author={Chen, Ziyang and Geng, Daniel and Owens, Andrew},
  journal={Advances in Neural Information Processing Systems},
  volume={37},
  pages={85045--85073},
  year={2024}
}

@inproceedings{sheffer2023hear,
  title={I hear your true colors: Image guided audio generation},
  author={Sheffer, Roy and Adi, Yossi},
  booktitle={ICASSP 2023-2023 IEEE International Conference on Acoustics, Speech and Signal Processing (ICASSP)},
  pages={1--5},
  year={2023},
  organization={IEEE}
}

@inproceedings{iashin2021taming,
  title={Taming Visually Guided Sound Generation},
  author={Iashin, Vladimir and Rahtu, Esa},
  booktitle={British Machine Vision Conference},
  year={2021},
  organization={BMVA Press}
}

@article{luo2023diff,
  title={Diff-foley: Synchronized video-to-audio synthesis with latent diffusion models},
  author={Luo, Simian and Yan, Chuanhao and Hu, Chenxu and Zhao, Hang},
  journal={Advances in Neural Information Processing Systems},
  volume={36},
  pages={48855--48876},
  year={2023}
}

@inproceedings{xing2024seeing,
  title={Seeing and hearing: Open-domain visual-audio generation with diffusion latent aligners},
  author={Xing, Yazhou and He, Yingqing and Tian, Zeyue and Wang, Xintao and Chen, Qifeng},
  booktitle={Proceedings of the IEEE/CVF Conference on Computer Vision and Pattern Recognition},
  pages={7151--7161},
  year={2024}
}

@inproceedings{wang2024v2a,
  title={V2a-mapper: A lightweight solution for vision-to-audio generation by connecting foundation models},
  author={Wang, Heng and Ma, Jianbo and Pascual, Santiago and Cartwright, Richard and Cai, Weidong},
  booktitle={Proceedings of the AAAI Conference on Artificial Intelligence},
  volume={38},
  number={14},
  pages={15492--15501},
  year={2024}
}

@inproceedings{xie2024sonicvisionlm,
  title={Sonicvisionlm: Playing sound with vision language models},
  author={Xie, Zhifeng and Yu, Shengye and He, Qile and Li, Mengtian},
  booktitle={Proceedings of the IEEE/CVF Conference on Computer Vision and Pattern Recognition},
  pages={26866--26875},
  year={2024}
}

@inproceedings{jeong2025read,
  title={Read, watch and scream! sound generation from text and video},
  author={Jeong, Yujin and Kim, Yunji and Chun, Sanghyuk and Lee, Jiyoung},
  booktitle={Proceedings of the AAAI Conference on Artificial Intelligence},
  volume={39},
  number={17},
  pages={17590--17598},
  year={2025}
}

@article{zhang2026foleycrafter,
  title={Foleycrafter: Bring silent videos to life with lifelike and synchronized sounds},
  author={Zhang, Yiming and Gu, Yicheng and Zeng, Yanhong and Xing, Zhening and Wang, Yuancheng and Wu, Zhizheng and Liu, Bin and Chen, Kai},
  journal={International Journal of Computer Vision},
  volume={134},
  number={1},
  pages={46},
  year={2026},
  publisher={Springer}
}

@article{tian2025audiox,
  title={Audiox: Diffusion transformer for anything-to-audio generation},
  author={Tian, Zeyue and Jin, Yizhu and Liu, Zhaoyang and Yuan, Ruibin and Tan, Xu and Chen, Qifeng and Xue, Wei and Guo, Yike},
  journal={arXiv preprint arXiv:2503.10522},
  year={2025}
}

@article{wang2025audiogen,
  title={Audiogen-omni: A unified multimodal diffusion transformer for video-synchronized audio, speech, and song generation},
  author={Wang, Le and Wang, Jun and Qiang, Chunyu and Deng, Feng and Zhang, Chen and Zhang, Di and Gai, Kun},
  journal={arXiv preprint arXiv:2508.00733},
  year={2025}
}

@article{wang2025kling,
  title={Kling-foley: Multimodal diffusion transformer for high-quality video-to-audio generation},
  author={Wang, Jun and Zeng, Xijuan and Qiang, Chunyu and Chen, Ruilong and Wang, Shiyao and Wang, Le and Zhou, Wangjing and Cai, Pengfei and Zhao, Jiahui and Li, Nan and others},
  journal={arXiv preprint arXiv:2506.19774},
  year={2025}
}

@article{xu2025uniflow,
  title={Uniflow-audio: Unified flow matching for audio generation from omni-modalities},
  author={Xu, Xuenan and Mei, Jiahao and Zheng, Zihao and Tao, Ye and Xie, Zeyu and Zhang, Yaoyun and Liu, Haohe and Wu, Yuning and Yan, Ming and Wu, Wen and others},
  journal={arXiv preprint arXiv:2509.24391},
  year={2025}
}

@article{dai2026omni2sound,
  title={Omni2Sound: Towards Unified Video-Text-to-Audio Generation},
  author={Dai, Yusheng and Chen, Zehua and Jiang, Yuxuan and Gao, Baolong and Ke, Qiuhong and Zhu, Jun and Cai, Jianfei},
  journal={arXiv preprint arXiv:2601.02731},
  year={2026}
}

@article{salimans2016improved,
  title={Improved techniques for training gans},
  author={Salimans, Tim and Goodfellow, Ian and Zaremba, Wojciech and Cheung, Vicki and Radford, Alec and Chen, Xi},
  journal={Advances in neural information processing systems},
  volume={29},
  year={2016}
}

@inproceedings{chen2020vggsound,
  title={Vggsound: A large-scale audio-visual dataset},
  author={Chen, Honglie and Xie, Weidi and Vedaldi, Andrea and Zisserman, Andrew},
  booktitle={ICASSP 2020-2020 IEEE International Conference on Acoustics, Speech and Signal Processing (ICASSP)},
  pages={721--725},
  year={2020},
  organization={IEEE}
}

@inproceedings{zverev2025vggsounder,
  title={Vggsounder: Audio-visual evaluations for foundation models},
  author={Zverev, Daniil and Wiedemer, Thadd{\"a}us and Prabhu, Ameya and Bethge, Matthias and Brendel, Wieland and Koepke, A},
  booktitle={Proceedings of the IEEE/CVF International Conference on Computer Vision},
  pages={1027--1037},
  year={2025}
}

@article{wang2024frieren,
  title={Frieren: Efficient video-to-audio generation network with rectified flow matching},
  author={Wang, Yongqi and Guo, Wenxiang and Huang, Rongjie and Huang, Jiawei and Wang, Zehan and You, Fuming and Li, Ruiqi and Zhao, Zhou},
  journal={Advances in neural information processing systems},
  volume={37},
  pages={128118--128138},
  year={2024}
}

@inproceedings{liuthinksound,
  title={ThinkSound: Chain-of-Thought Reasoning in Multimodal LLMs for Audio Generation and Editing},
  author={Liu, Huadai and Luo, Kaicheng and Wang, Jialei and Wang, Wen and Chen, Qian and Zhao, Zhou and Xue, Wei},
  booktitle={The Thirty-ninth Annual Conference on Neural Information Processing Systems}
}

@article{comanici2025gemini,
  title={Gemini 2.5: Pushing the frontier with advanced reasoning, multimodality, long context, and next generation agentic capabilities},
  author={Comanici, Gheorghe and Bieber, Eric and Schaekermann, Mike and Pasupat, Ice and Sachdeva, Noveen and Dhillon, Inderjit and Blistein, Marcel and Ram, Ori and Zhang, Dan and Rosen, Evan and others},
  journal={arXiv preprint arXiv:2507.06261},
  year={2025}
}

@article{hurst2024gpt,
  title={Gpt-4o system card},
  author={Hurst, Aaron and Lerer, Adam and Goucher, Adam P and Perelman, Adam and Ramesh, Aditya and Clark, Aidan and Ostrow, AJ and Welihinda, Akila and Hayes, Alan and Radford, Alec and others},
  journal={arXiv preprint arXiv:2410.21276},
  year={2024}
}

@inproceedings{sun2024video,
  title={video-SALMONN: speech-enhanced audio-visual large language models},
  author={Sun, Guangzhi and Yu, Wenyi and Tang, Changli and Chen, Xianzhao and Tan, Tian and Li, Wei and Lu, Lu and Ma, Zejun and Wang, Yuxuan and Zhang, Chao},
  booktitle={Proceedings of the 41st International Conference on Machine Learning},
  pages={47198--47217},
  year={2024}
}

@inproceedings{zhang2023video,
  title={Video-llama: An instruction-tuned audio-visual language model for video understanding},
  author={Zhang, Hang and Li, Xin and Bing, Lidong},
  booktitle={Proceedings of the 2023 conference on empirical methods in natural language processing: system demonstrations},
  pages={543--553},
  year={2023}
}

@article{liang2026omni,
  title={Omni-Judge: Can Omni-LLMs Serve as Human-Aligned Judges for Text-Conditioned Audio-Video Generation?},
  author={Liang, Susan and Huang, Chao and Bellos, Filippos and Tang, Yolo Yunlong and Shen, Qianxiang and Bi, Jing and Song, Luchuan and Zhang, Zeliang and Corso, Jason and Xu, Chenliang},
  journal={arXiv preprint arXiv:2602.01623},
  year={2026}
}

@inproceedings{zhou2018visual,
  title={Visual to sound: Generating natural sound for videos in the wild},
  author={Zhou, Yipin and Wang, Zhaowen and Fang, Chen and Bui, Trung and Berg, Tamara L},
  booktitle={Proceedings of the IEEE conference on computer vision and pattern recognition},
  pages={3550--3558},
  year={2018}
}

@inproceedings{radford2021learning,
  title={Learning transferable visual models from natural language supervision},
  author={Radford, Alec and Kim, Jong Wook and Hallacy, Chris and Ramesh, Aditya and Goh, Gabriel and Agarwal, Sandhini and Sastry, Girish and Askell, Amanda and Mishkin, Pamela and Clark, Jack and others},
  booktitle={International conference on machine learning},
  pages={8748--8763},
  year={2021},
  organization={PmLR}
}

@inproceedings{liu2023audioldm,
  title={AudioLDM: Text-to-Audio Generation with Latent Diffusion Models},
  author={Liu, Haohe and Chen, Zehua and Yuan, Yi and Mei, Xinhao and Liu, Xubo and Mandic, Danilo and Wang, Wenwu and Plumbley, Mark},
  booktitle={Proceedings of the 40th International Conference on Machine Learning, PMLR 2023},
  volume={202},
  pages={21450--21474},
  year={2023},
  organization={International Machine Learning Society (IMLS)}
}

@inproceedings{viertola2025temporally,
  title={Temporally aligned audio for video with autoregression},
  author={Viertola, Ilpo and Iashin, Vladimir and Rahtu, Esa},
  booktitle={ICASSP 2025-2025 IEEE International Conference on Acoustics, Speech and Signal Processing (ICASSP)},
  pages={1--5},
  year={2025},
  organization={IEEE}
}

@article{akbari2021vatt,
  title={Vatt: Transformers for multimodal self-supervised learning from raw video, audio and text},
  author={Akbari, Hassan and Yuan, Liangzhe and Qian, Rui and Chuang, Wei-Hong and Chang, Shih-Fu and Cui, Yin and Gong, Boqing},
  journal={Advances in neural information processing systems},
  volume={34},
  pages={24206--24221},
  year={2021}
}

@inproceedings{chen2025video,
  title={Video-guided foley sound generation with multimodal controls},
  author={Chen, Ziyang and Seetharaman, Prem and Russell, Bryan and Nieto, Oriol and Bourgin, David and Owens, Andrew and Salamon, Justin},
  booktitle={Proceedings of the Computer Vision and Pattern Recognition Conference},
  pages={18770--18781},
  year={2025}
}

@article{shan2025hunyuanvideo,
  title={Hunyuanvideo-foley: Multimodal diffusion with representation alignment for high-fidelity foley audio generation},
  author={Shan, Sizhe and Li, Qiulin and Cui, Yutao and Yang, Miles and Wang, Yuehai and Yang, Qun and Zhou, Jin and Zhong, Zhao},
  journal={arXiv preprint arXiv:2508.16930},
  year={2025}
}

@inproceedings{huang2024vbench,
  title={Vbench: Comprehensive benchmark suite for video generative models},
  author={Huang, Ziqi and He, Yinan and Yu, Jiashuo and Zhang, Fan and Si, Chenyang and Jiang, Yuming and Zhang, Yuanhan and Wu, Tianxing and Jin, Qingyang and Chanpaisit, Nattapol and others},
  booktitle={Proceedings of the IEEE/CVF Conference on Computer Vision and Pattern Recognition},
  pages={21807--21818},
  year={2024}
}

@inproceedings{liu2024evalcrafter,
  title={Evalcrafter: Benchmarking and evaluating large video generation models},
  author={Liu, Yaofang and Cun, Xiaodong and Liu, Xuebo and Wang, Xintao and Zhang, Yong and Chen, Haoxin and Liu, Yang and Zeng, Tieyong and Chan, Raymond and Shan, Ying},
  booktitle={Proceedings of the IEEE/CVF conference on computer vision and pattern recognition},
  pages={22139--22149},
  year={2024}
}

@inproceedings{wang2025t2a,
  title={T2A-Feedback: Improving Basic Capabilities of Text-to-Audio Generation via Fine-grained AI Feedback},
  author={Wang, Zehan and Lei, Ke and Zhu, Chen and Huang, Jiawei and Zhou, Sashuai and Liu, Luping and Cheng, Xize and Ji, Shengpeng and Ye, Zhenhui and Jin, Tao and others},
  booktitle={Proceedings of the 63rd Annual Meeting of the Association for Computational Linguistics},
  pages={23535--23547},
  year={2025}
}

@inproceedings{wang2026tta,
  title={Tta-bench: A comprehensive benchmark for evaluating text-to-audio models},
  author={Wang, Hui and Liu, Cheng and Chen, Junyang and Liu, Haoze and Jia, Yuhang and Zhao, Shiwan and Zhou, Jiaming and Sun, Haoqin and Bu, Hui and Qin, Yong},
  booktitle={Proceedings of the AAAI Conference on Artificial Intelligence},
  volume={40},
  number={39},
  pages={33512--33520},
  year={2026}
}

@article{hua2025vabench,
  title={Vabench: A comprehensive benchmark for audio-video generation},
  author={Hua, Daili and Wang, Xizhi and Zeng, Bohan and Huang, Xinyi and Liang, Hao and Niu, Junbo and Chen, Xinlong and Xu, Quanqing and Zhang, Wentao},
  journal={arXiv preprint arXiv:2512.09299},
  year={2025}
}

@article{cao2025t2av,
  title={T2AV-Compass: Towards Unified Evaluation for Text-to-Audio-Video Generation},
  author={Cao, Zhe and Wang, Tao and Wang, Jiaming and Wang, Yanghai and Zhang, Yuanxing and Chen, Jialu and Deng, Miao and Wang, Jiahao and Guo, Yubin and Liao, Chenxi and others},
  journal={arXiv preprint arXiv:2512.21094},
  year={2025}
}

@inproceedings{zhou2025harmonyset,
  title={Harmonyset: A comprehensive dataset for understanding video-music semantic alignment and temporal synchronization},
  author={Zhou, Zitang and Mei, Ke and Lu, Yu and Wang, Tianyi and Rao, Fengyun},
  booktitle={Proceedings of the Computer Vision and Pattern Recognition Conference},
  pages={3152--3162},
  year={2025}
}

@article{ephrat2018looking,
  title={Looking to listen at the cocktail party: a speaker-independent audio-visual model for speech separation},
  author={Ephrat, Ariel and Mosseri, Inbar and Lang, Oran and Dekel, Tali and Wilson, Kevin and Hassidim, Avinatan and Freeman, William T and Rubinstein, Michael},
  journal={ACM Transactions on Graphics (TOG)},
  volume={37},
  number={4},
  pages={1--11},
  year={2018},
  publisher={ACM New York, NY, USA}
}

@article{montesinos2021cappella,
  title={A cappella: Audio-visual singing voice separation},
  author={Montesinos, Juan F and Kadandale, Venkatesh S and Haro, Gloria},
  journal={arXiv preprint arXiv:2104.09946},
  year={2021}
}

@article{bai2025qwen3,
  title={Qwen3-vl technical report},
  author={Bai, Shuai and Cai, Yuxuan and Chen, Ruizhe and Chen, Keqin and Chen, Xionghui and Cheng, Zesen and Deng, Lianghao and Ding, Wei and Gao, Chang and Ge, Chunjiang and others},
  journal={arXiv preprint arXiv:2511.21631},
  year={2025}
}

@article{taal2011algorithm,
  title={An algorithm for intelligibility prediction of time--frequency weighted noisy speech},
  author={Taal, Cees H and Hendriks, Richard C and Heusdens, Richard and Jensen, Jesper},
  journal={IEEE Transactions on audio, speech, and language processing},
  volume={19},
  number={7},
  pages={2125--2136},
  year={2011},
  publisher={IEEE}
}

@inproceedings{zuo2025gvmgen,
  title={Gvmgen: A general video-to-music generation model with hierarchical attentions},
  author={Zuo, Heda and You, Weitao and Wu, Junxian and Ren, Shihong and Chen, Pei and Zhou, Mingxu and Lu, Yujia and Sun, Lingyun},
  booktitle={Proceedings of the AAAI Conference on Artificial Intelligence},
  volume={39},
  number={21},
  pages={23099--23107},
  year={2025}
}

@inproceedings{zhang2025sonique,
  title={Sonique: Video background music generation using unpaired audio-visual data},
  author={Zhang, Liqian and Fuentes, Magdalena},
  booktitle={ICASSP 2025-2025 IEEE International Conference on Acoustics, Speech and Signal Processing (ICASSP)},
  pages={1--5},
  year={2025},
  organization={IEEE}
}

@inproceedings{tian2025vidmuse,
  title={Vidmuse: A simple video-to-music generation framework with long-short-term modeling},
  author={Tian, Zeyue and Liu, Zhaoyang and Yuan, Ruibin and Pan, Jiahao and Liu, Qifeng and Tan, Xu and Chen, Qifeng and Xue, Wei and Guo, Yike},
  booktitle={Proceedings of the Computer Vision and Pattern Recognition Conference},
  pages={18782--18793},
  year={2025}
}

@inproceedings{di2021video,
  title={Video background music generation with controllable music transformer},
  author={Di, Shangzhe and Jiang, Zeren and Liu, Si and Wang, Zhaokai and Zhu, Leyan and He, Zexin and Liu, Hongming and Yan, Shuicheng},
  booktitle={Proceedings of the 29th ACM International Conference on Multimedia},
  pages={2037--2045},
  year={2021}
}

@article{berthe2026s,
  title={S-VoCAL: A Dataset and Evaluation Framework for Inferring Speaking Voice Character Attributes in Literature},
  author={Berthe-Pardo, Abigail and Michel, Gaspard and Epure, Elena V and Cerisara, Christophe},
  journal={arXiv preprint arXiv:2603.00958},
  year={2026}
}

@inproceedings{gemmeke2017audio,
  title={Audio set: An ontology and human-labeled dataset for audio events},
  author={Gemmeke, Jort F and Ellis, Daniel PW and Freedman, Dylan and Jansen, Aren and Lawrence, Wade and Moore, R Channing and Plakal, Manoj and Ritter, Marvin},
  booktitle={2017 IEEE international conference on acoustics, speech and signal processing (ICASSP)},
  pages={776--780},
  year={2017},
  organization={IEEE}
}


\clearpage
\appendix

\section{VidAudio-Bench Construction}
To support reliable evaluation of V2A and VT2A systems, we construct subsets with clear audio-visual grounding, temporally complete events, and minimal confounding factors such as background music, narration, and static imagery. In this section, we describe the subset construction procedure of VidAudio-Bench and provide additional details of the V2A and VT2A settings.

\subsection{Subset Construction and Statistics}

\noindent\textbf{SFX and Instrument Performance subsets.}
The \textit{SFX} and \textit{Instrument Performance} subsets are constructed from VGGSounder.

We first enforce a strict \textit{audio-visual consistency} criterion by retaining only labels whose \textit{modality} is annotated as \textit{AV}, indicating that the sound source is visually observable in the video. We further exclude videos annotated with \texttt{static\_image}, \texttt{background\_music}, or \texttt{voice\_over}, thereby removing samples containing static imagery, background music, or voice-overs. From the remaining candidates, we retain only videos containing either a single dominant sound event or a clearly identifiable primary sound source.

Based on semantic attributes, we group the 300 candidate labels into five categories and manually verify the grouping: \textit{sound effects}, \textit{instrument}, \textit{singing}, \textit{speech}, and \textit{others}. Among them, 225 labels belong to \textit{sound effects}, 55 to \textit{instrument}, 6 to \textit{singing}, 3 to \textit{speech}, and 11 to \textit{others}.

For the \textit{SFX} subset, we sample approximately 1--2 videos from each label in the \textit{sound effects} category, resulting in 400 videos in total. For the \textit{Instrument Performance} subset, we sample approximately 3--4 videos from each label in the \textit{instrument} category, resulting in 191 videos in total.

\begin{figure}[h]
  \centering
  \includegraphics[width=0.85\linewidth]{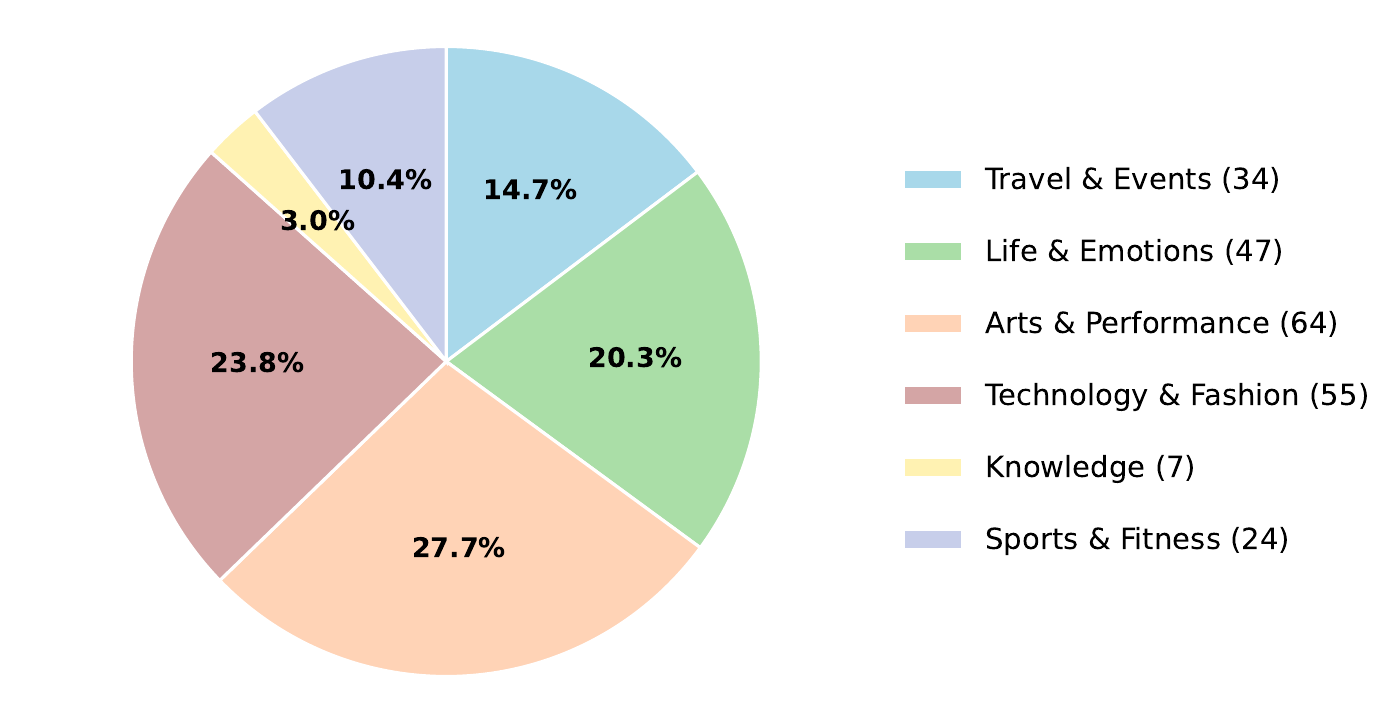}
  \vspace{-4pt}
  \caption{Content category distribution of the \textit{BGM} subset.}
  \label{musicB}
\end{figure}

\noindent\textbf{BGM subset.}
The \textit{BGM} subset is constructed from the test split of HarmonySet. We first retain videos with durations in the range of 9--11 seconds, and then randomly sample 231 videos from the filtered set. To preserve semantic and temporal completeness, each selected video is further adjusted by slight padding or trimming to 10 seconds when necessary. The category distribution of the resulting subset is shown in Figure~\ref{musicB}.

\vspace{-6pt}
\vspace{-6pt}
\noindent\textbf{Speech subset.}
The \textit{Speech} subset is constructed from the test split of AVSpeech. We first retain videos with durations of 9--11 seconds and standardize them to a uniform 10 seconds by padding or trimming. We then manually remove samples in which the speaker is visually too small for reliable lip-motion perception, as such videos usually contain unclear or barely visible lip movements. After this filtering process, 412 videos are retained in the final \textit{Speech} subset.

\noindent\textbf{Singing subset.}
The \textit{Singing} subset is constructed from the test split of Acappella. We first retain English singing videos and then segment the original videos into 10-second clips. From these clips, we randomly sample 400 videos to form the final subset.

\subsection{V2A and VT2A Settings}
In this section, we provide additional details on the prompt design used in our benchmark. Appendix~\ref{A.2.1} describes the instruction templates for the two evaluation settings, while Appendix~\ref{A.2.2} presents the prompts used for visual captioning with Qwen3-VL.

\subsubsection{Instruction Design for V2A and VT2A}
\label{A.2.1}

Table~\ref{tab:instruction_templates} summarizes the instruction templates used in the V2A and VT2A settings. The prompting strategy is designed according to the capabilities of different models. For models that support negative prompting (e.g., AudioX, HunyuanVideo, and MMAudio), we include negative instructions to explicitly constrain the generation and better guide the output toward the target audio category. For models that do not support negative conditioning (e.g., FoleyCrafter, ReSound, ThinkSound, UniFlow-Audio, and Kling), we use only positive instructions. This design choice is motivated by the observation that explicitly mentioning undesired audio concepts may unintentionally bias the generation process, thereby increasing the risk of irrelevant or hallucinated audio content. In the VT2A setting, we use Qwen3-VL for visual caption generation, and Qwen3.5 only to rewrite the VT2A prompts into more natural text descriptions. 

\begin{table}[htbp]
\centering
\caption{Predefined positive and negative instruction templates used for the V2A and VT2A settings across four audio categories in \textit{VidAudio-Bench}, where the Music category is further divided into Instrument Performance and BGM.}
\label{tab:instruction_templates}
\footnotesize
\setlength{\tabcolsep}{3.5pt}
\renewcommand{\arraystretch}{1.15}
\begin{tabularx}{\columnwidth}{l l X X}
\toprule
\textbf{Domain} & \textbf{Task} & \textbf{Positive Instruction (Input Text)} & \textbf{Negative Instruction} \\ \midrule

\multirow{2}{*}{SFX} & V2A & Realistic foley sound synchronized with the video. & music, background music, speech, singing \\ \cmidrule{2-4}
 & VT2A & Realistic foley sound of {\color{blue}\{caption\}}. & music, background music, speech, singing \\ \midrule

\multirow{2}{*}{\makecell[l]{Music-\\Instrument\\Performance}} & V2A & Musical instrument performance synchronized with the video. & speech, singing, human voice \\ \cmidrule{2-4}
& VT2A & Instrument performance of {\color{blue}\{caption\}}. & speech, singing, human voice \\ \midrule

\multirow{2}{*}{\makecell[l]{Music-\\BGM}} & V2A & Background music matching the video scene. & speech, sound effects, foley \\ \cmidrule{2-4}
 & VT2A & Background music fitting the scene: {\color{blue}\{caption\}}. & speech, sound effects, foley \\ \midrule

\multirow{2}{*}{Speech} & V2A & Human speech synchronized with the video. & background music, singing, noise, sound effects \\ \cmidrule{2-4}
 & VT2A & Speech by {\color{blue}\{caption\}}. &  background music, singing, noise, sound effects \\ \midrule

\multirow{2}{*}{Singing} & V2A & A cappella singing voice synchronized with the video. & speech, talking, instrumental only, heavy accompaniment \\ \cmidrule{2-4}
 & VT2A & Singing performance by {\color{blue}\{caption\}}. & speech, talking, instrumental only, heavy accompaniment \\ 
\bottomrule
\end{tabularx}
\end{table}
\clearpage
\subsubsection{Visual Captioning Prompt Design}
\label{A.2.2}
\begin{figure}[h]
  \centering
  \includegraphics[width=0.95\linewidth]{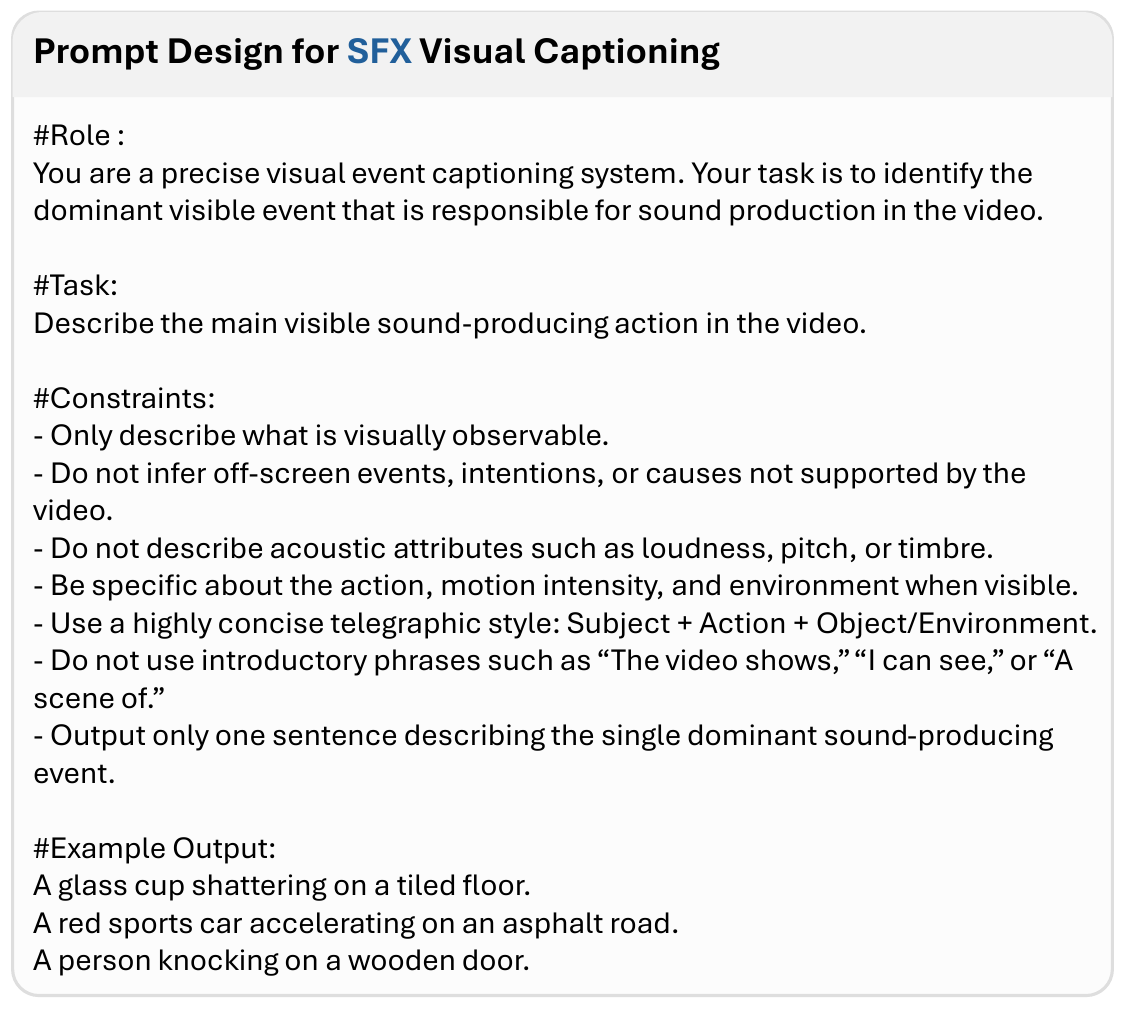}
  \caption{Prompt design for \textit{SFX} visual captioning}
  \label{fig:8}
\end{figure}

\begin{figure}[h]
  \centering
    \vspace{-6pt}
  \includegraphics[width=0.9\linewidth]{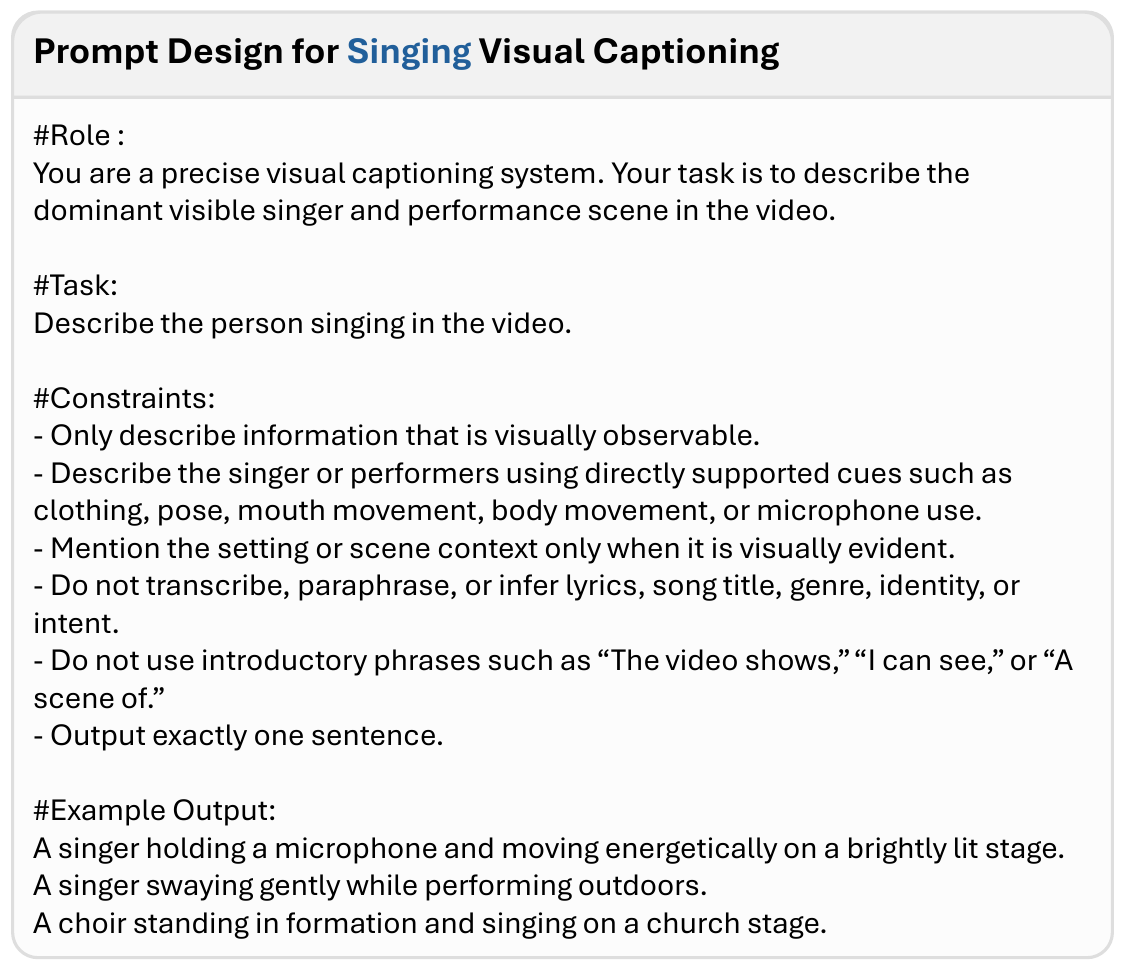}
  \caption{Prompt design for \textit{Singing} visual captioning}
   \label{fig:12}
\end{figure}

To generate visual captions that are better matched to the semantic characteristics of different audio categories, we adopt a category-specific prompting strategy for the Vision-Language Model (as shown in Figures~\ref{fig:8}--\ref{fig:12}). Specifically, different prompts are designed for sound effects, music, speech, and singing videos, with the music category further divided into instrument-performance and background-music scenarios. By explicitly adapting the instruction focus to each category, the model is encouraged to attend to the most relevant visual cues and avoid category-irrelevant or speculative descriptions, resulting in captions that are more accurate, informative, and semantically aligned with the video content.

\vspace{-10pt}

\begin{figure}[h]
  \centering
  \includegraphics[width=0.95\linewidth]{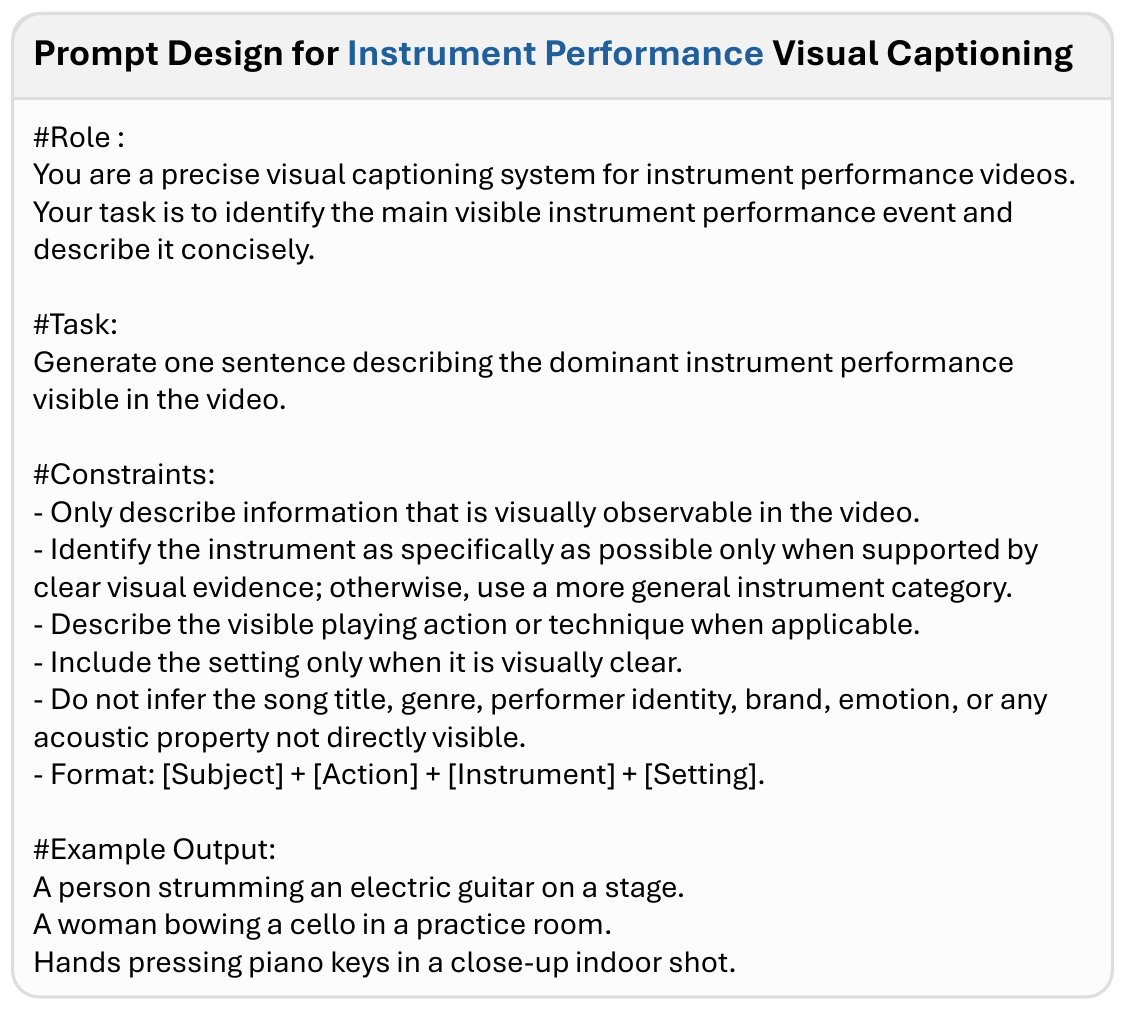}
  \caption{Prompt design for \textit{Instrument Performance} visual captioning}
\end{figure}
\vspace{-10pt}

\begin{figure}[h]
  \centering
  \includegraphics[width=0.9\linewidth]{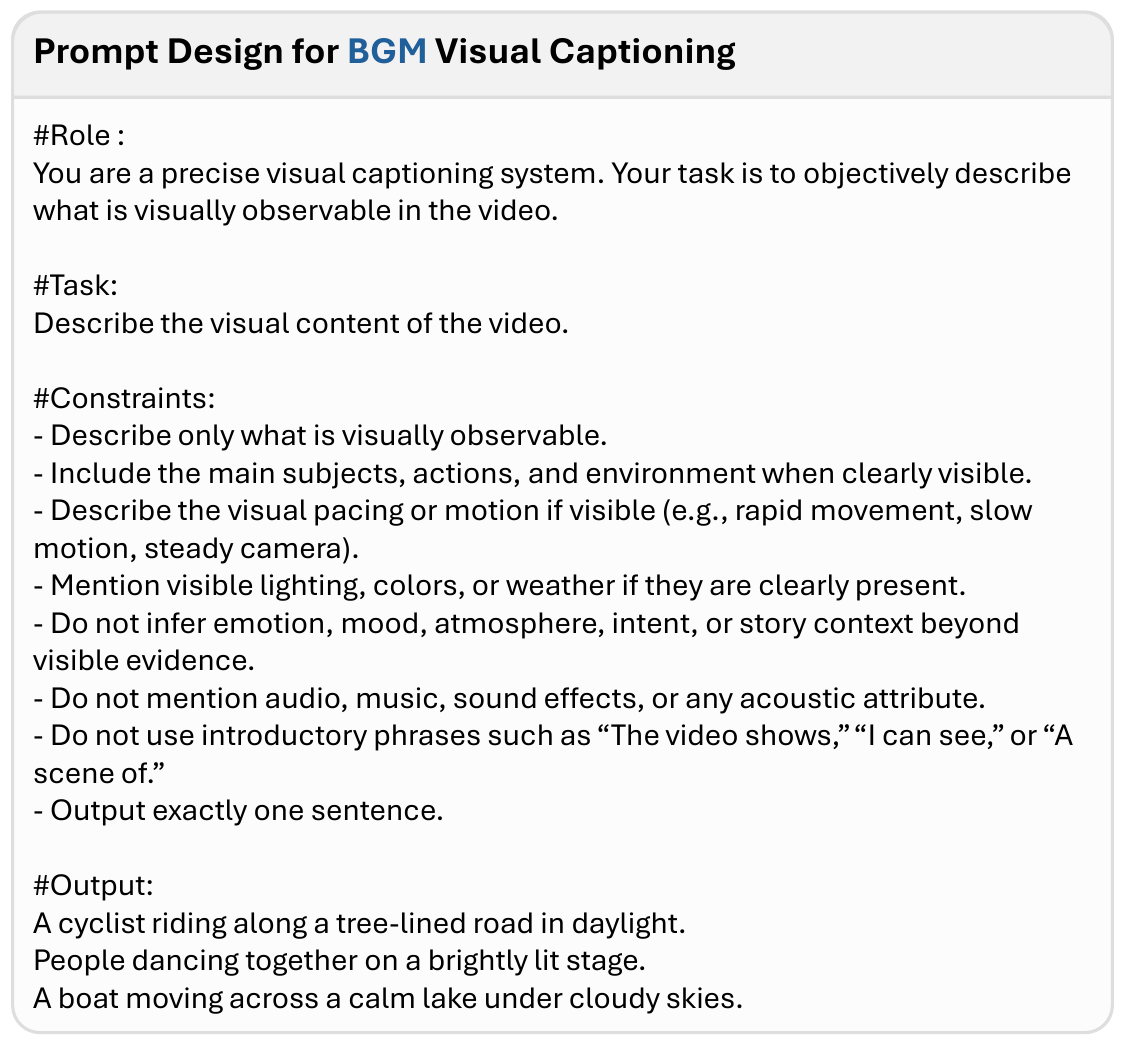}
  \caption{Prompt design for \textit{BGM} visual captioning}
\end{figure}

\vspace{-10pt}
\begin{figure}[h]
  \centering
  \vspace{-6pt}
  \includegraphics[width=0.9\linewidth]{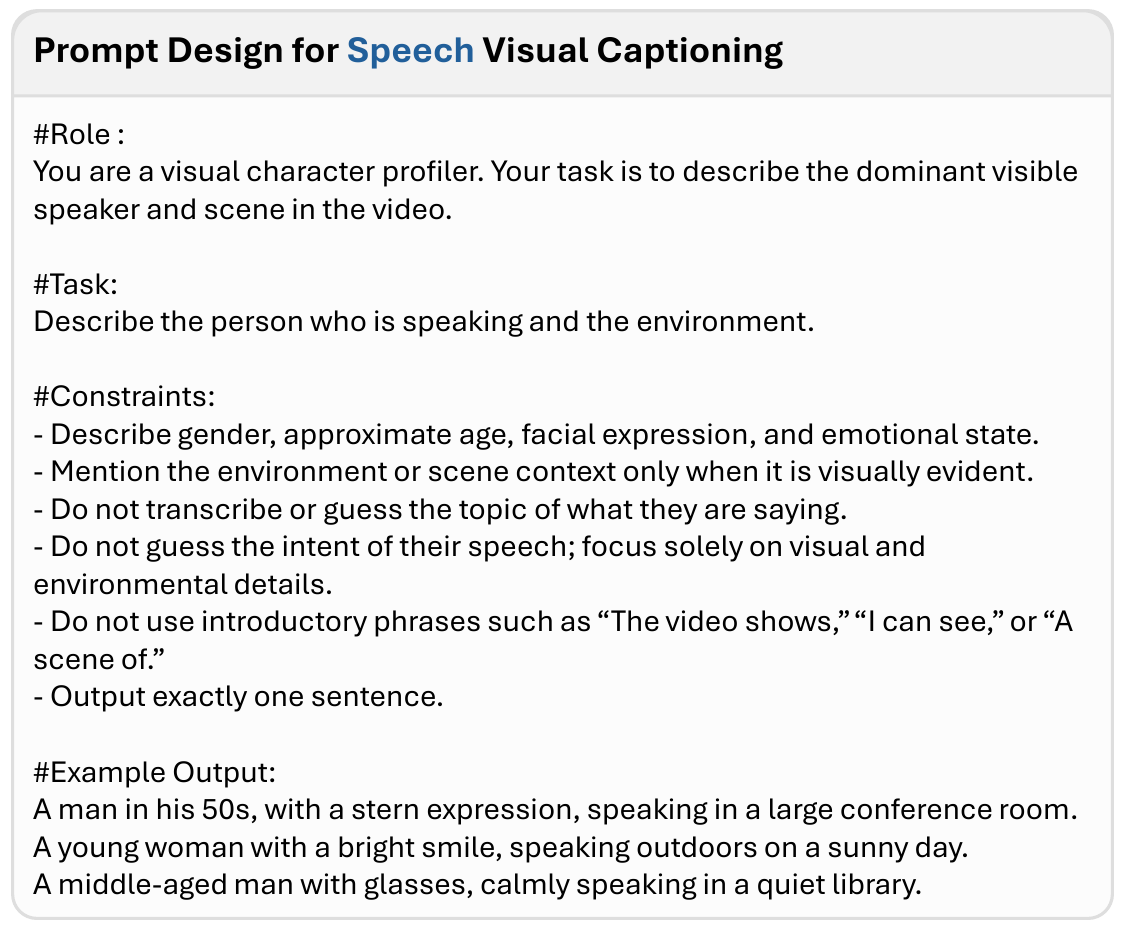}
  \caption{Prompt design for \textit{Speech} visual captioning}
\end{figure}
\vspace{-10pt}

\clearpage

\section{Evaluation Metrics}
In this section, we present the prompts used for the three evaluation methods involving MLLM-as-a-Judge discussed in the main text.

\subsection{Identity Consistency}
\begin{figure}[!t]
  \centering
  \includegraphics[width=0.95\linewidth]{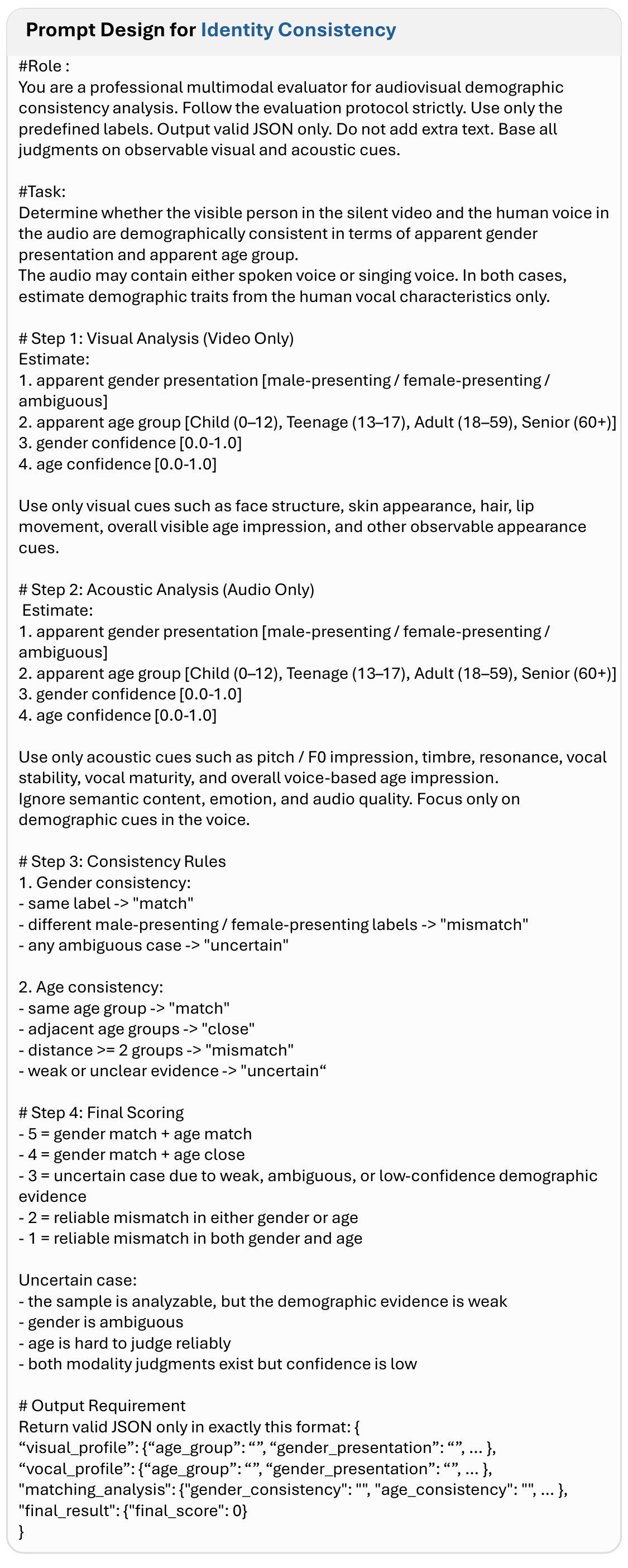}
  \caption{Prompt design for \textit{Identity Consistency}.}
  \label{fig:13}
\end{figure}

To evaluate cross-modal \textit{Identity Consistency}, we adopt a structured protocol that examines whether the visible person in the video and the human voice in the audio are demographically aligned. Specifically, the evaluation is conducted from two complementary aspects, i.e., apparent gender presentation and apparent age group. The framework first analyzes visual cues from the silent video and acoustic cues from the audio independently to construct the visual and vocal demographic profiles. It then compares the two profiles according to predefined consistency rules and assigns a final score based on the degree of agreement between modalities. This design enables a systematic assessment of whether the generated voice matches the on-screen person at the demographic level, while reducing interference from other factors such as semantic content, emotion, or audio quality. The detailed prompt used for this evaluation is illustrated in Figure~\ref{fig:13}.

\subsection{Affective Alignment}
\begin{figure}[!t]
  \centering
  \includegraphics[width=0.95\linewidth]{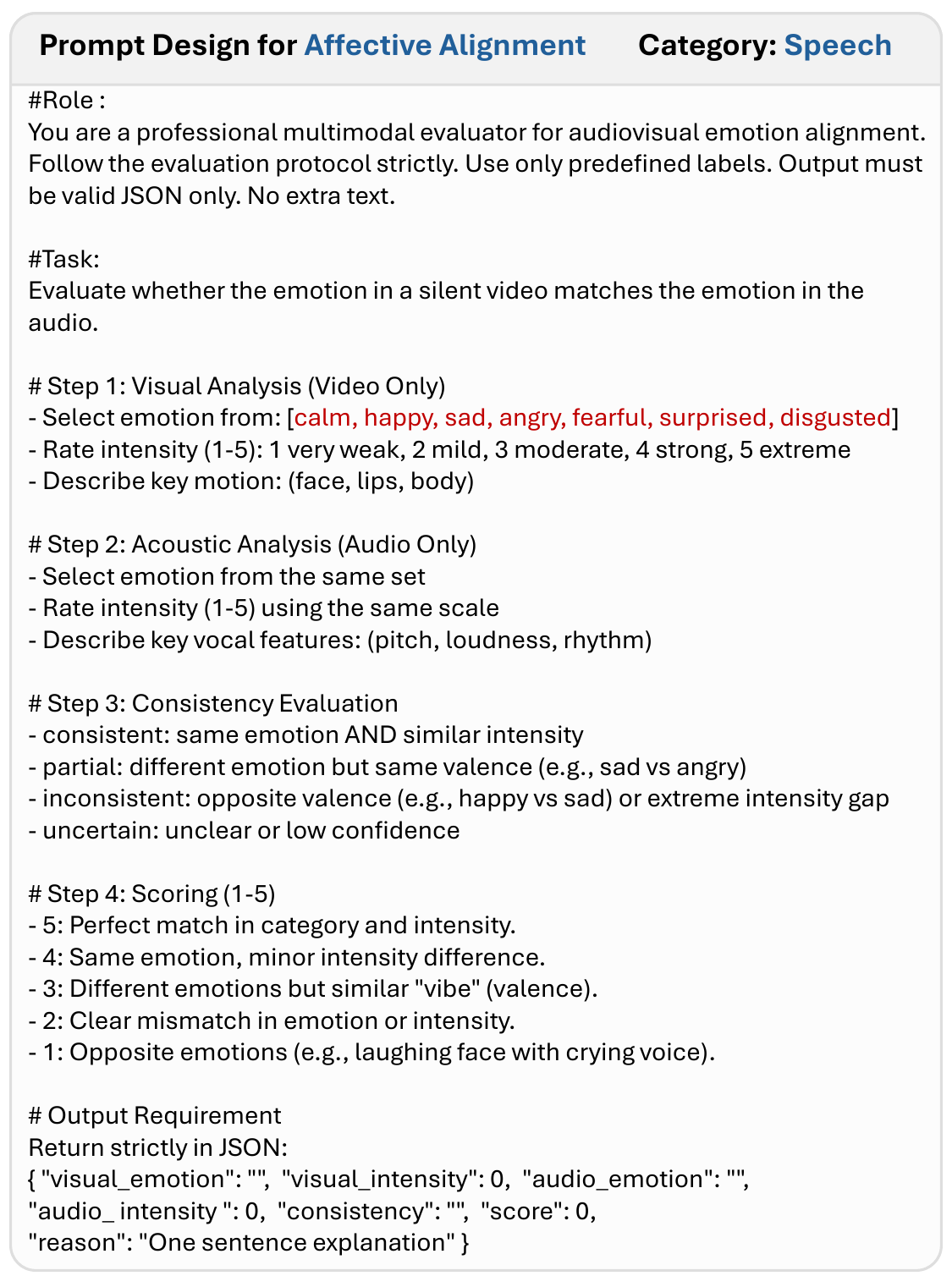}
  \caption{Prompt design for \textit{Affective Alignment} on \textit{Speech}.}
  \label{fig:14}
\end{figure}

\begin{figure}[!t]
  \centering
  \includegraphics[width=0.95\linewidth]{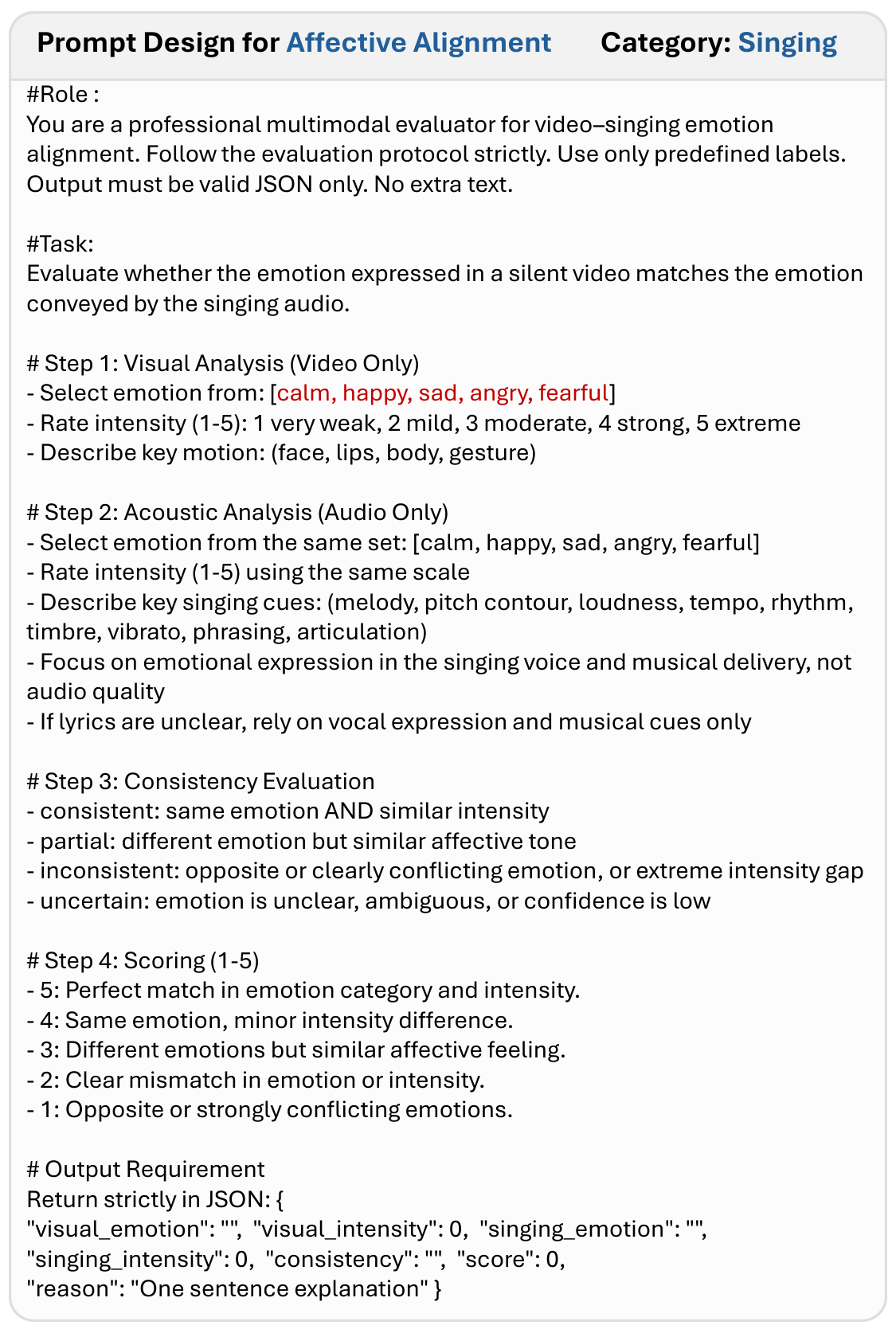}
  \caption{Prompt design for \textit{Affective Alignment} on \textit{Singing}.}
  \label{fig:15}
\end{figure}

Following the same evaluation framework as above, we further assess cross-modal \textit{Affective Alignment}, focusing on whether the emotion expressed in the silent video is consistent with that conveyed by the audio. In this setting, the evaluation considers two affective dimensions, namely emotion category and emotional intensity. Different emotion label sets are adopted for \textit{Speech} and \textit{Singing}. For speech, we use seven emotion categories, namely \textit{calm}, \textit{happy}, \textit{sad}, \textit{angry}, \textit{fearful}, \textit{surprised}, and \textit{disgusted}, as shown in Figure~\ref{fig:14}. For singing, we use five categories, namely \textit{calm}, \textit{happy}, \textit{sad}, \textit{angry}, and \textit{fearful}, as shown in Figure~\ref{fig:15}.

\subsection{Instruction Following}
\begin{figure*}[t]
  \centering
  \includegraphics[width=1\textwidth]{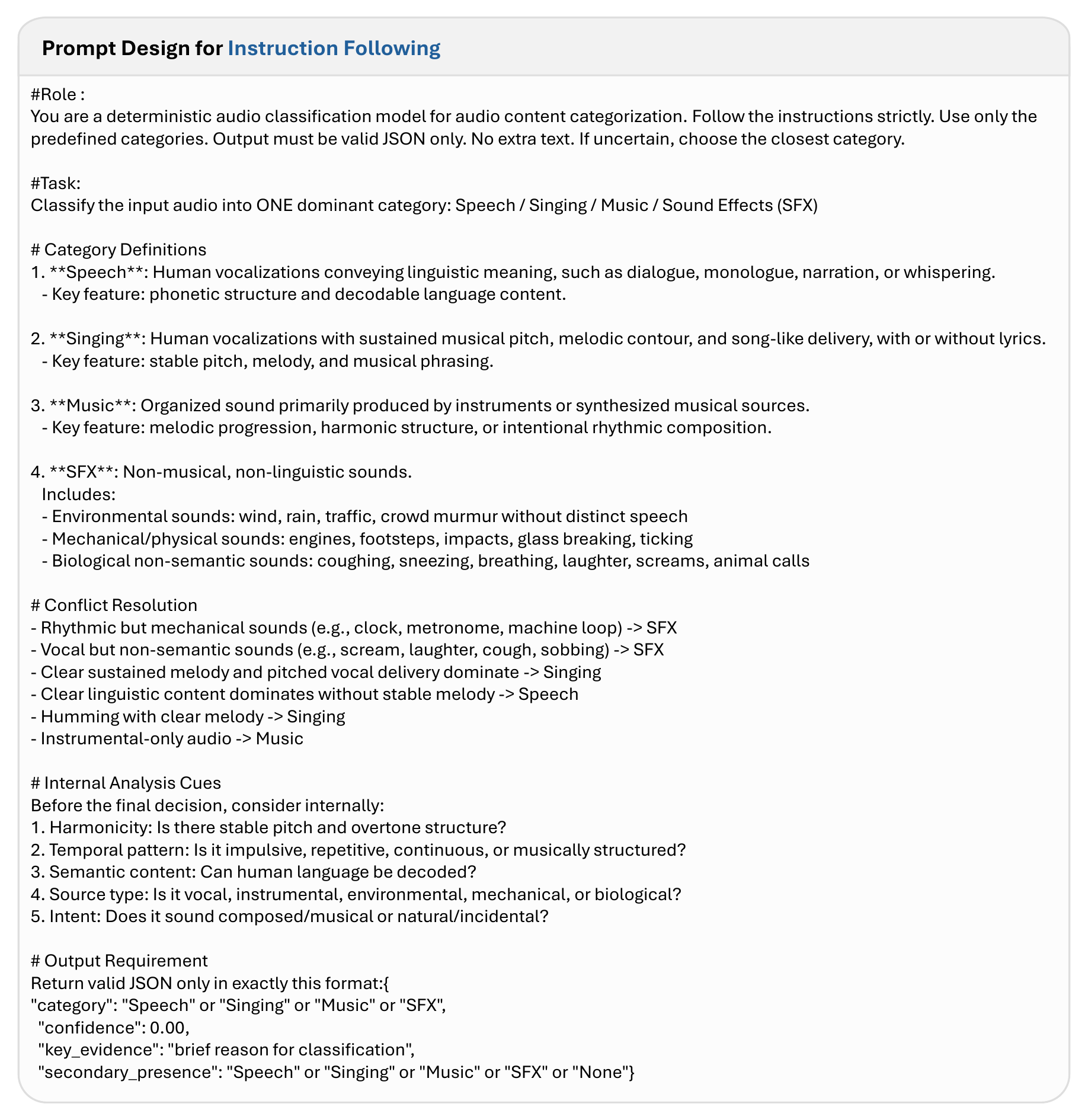}
  \caption{Prompt design for \textit{Instruction Following}.}
  \label{instructionfollow}
\end{figure*}

For instruction-following evaluation, we focus exclusively on category-level compliance of the generated audio. Specifically, for each generated sample, we assess only whether it belongs to the target audio category specified by the instruction, namely \textit{Speech}, \textit{Singing}, \textit{Music}, and \textit{Sound Effects (SFX)}. This evaluation aims to verify whether the model produces audio of the intended category, without considering finer-grained semantic correctness or perceptual quality. The detailed prompt used for this evaluation is illustrated in Figure~\ref{instructionfollow}.

\begin{table*}[!t]
\centering
\caption{Evaluation dimensions and task-specific criteria used in the human subjective study across four task categories. Each dimension was rated on a 5-point Likert scale (1--5), and the table summarizes the aspect emphasized for each task.}
\label{tab:human_rubrics}
\small
\begin{tabularx}{\linewidth}{l | X | X | X | X}
\toprule
\textbf{Dimension} & \textbf{SFX} & \textbf{Music} & \textbf{Speech} & \textbf{Singing} \\
\midrule
\textbf{Realism} & \multicolumn{4}{c}{\textbf{Fidelity:} Focuses on the clarity, fidelity, and absence of artifacts (e.g., noise, distortion) of the generated audio.} \\
\midrule
\textbf{Semantics} & \textbf{V-A Semantic-Corr}: Match between audio and visual sound sources or key events. & \textbf{V-A Semantic-Corr}: Match with specific instruments or overall visual atmosphere. & \textbf{Identity-Cons}: Match between the voice and the person's age and gender. & \textbf{Identity-Cons}: Match between the voice and the person's age and gender. \\
\midrule
\textbf{Synchronization} & \textbf{Temp-Sync}: Temporal alignment between visual actions and sound onsets. & \textbf{Temp-Sync}: Alignment with performance movements or rhythmic tempo. & \textbf{Lip-Sync}: Synchronization between mouth motion and speech articulation. & \textbf{Lip-Sync}: Synchronization between lip movements and melodic vocalization. \\
\midrule
\textbf{Instruction Following} & \textbf{Environmental/Event} sounds; minimal music or speech. & \textbf{Melodic/Harmonic} content; absence of distinct speech or dialogue. & \textbf{Intelligible human speech}; no dominant music or background SFX. & \textbf{Melodic vocal performance}; distinct from plain speech or pure music. \\
\bottomrule
\end{tabularx}
\end{table*}

\FloatBarrier

\begin{figure}[!t]
  \centering
  \includegraphics[width=0.85\linewidth]{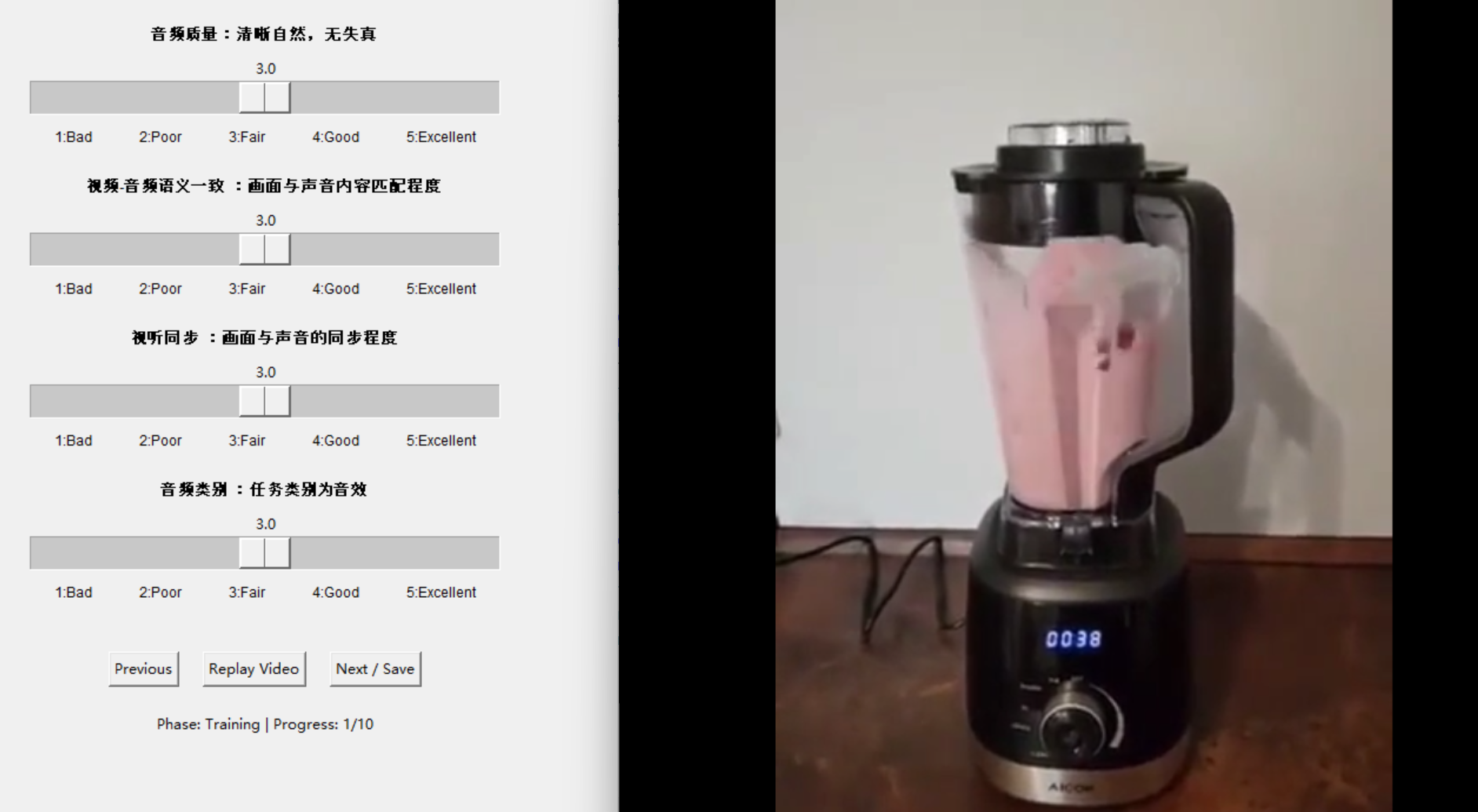}
  \caption{Annotation interface used in the human subjective study.}
  \label{fig:16}
\end{figure}

\begin{figure*}[t]
  \centering
  \includegraphics[width=1\textwidth]{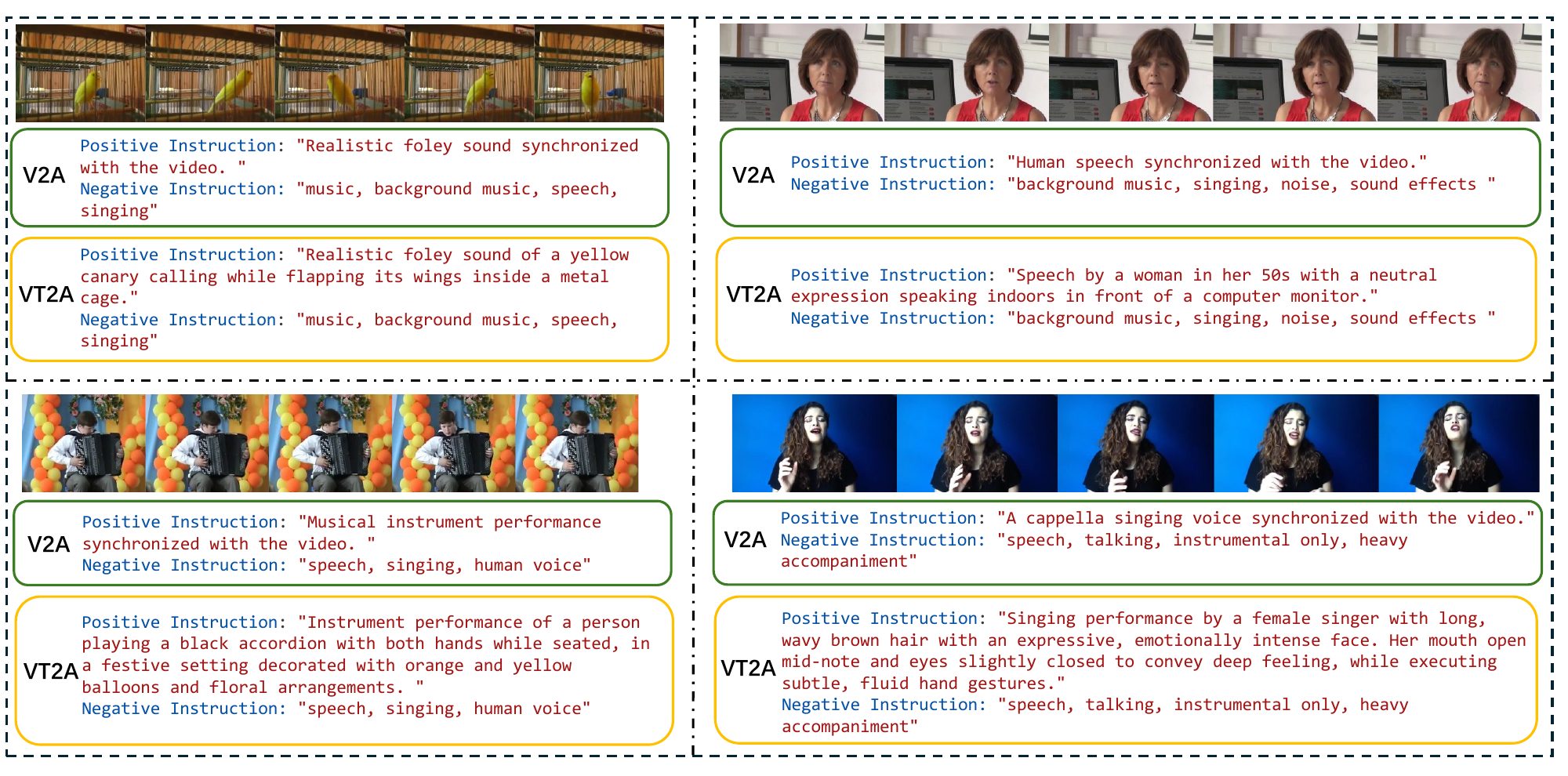}
  \caption{Examples of V2A and VT2A instruction prompts across four audio categories.}
  \label{examples}
\end{figure*}

\section{Human Subjective Study}
We recruited a total of 20 participants with normal vision and hearing. To ensure high-quality feedback, the participants were divided into four groups, with 5 experts assigned to each task category (SFX, Music, Speech, and Singing). The study was conducted in a controlled, noise-attenuated environment. All participants used professional-grade studio headphones to ensure they could discern subtle acoustic details and potential artifacts.

While certain metrics are common across all tasks, we designed specific dimensions to capture the unique nuances of different audio types. A Likert scale (1--5) was employed for all dimensions. The criteria for each task are summarized in Table~\ref{tab:human_rubrics}. A critical challenge in multimodal evaluation is category ambiguity (e.g., a video containing both ambient noise and a specific sound effect). To address this, we implemented a nuanced 1--5 scoring system for \textit{Instruction Following} rather than a binary ``yes/no'' choice. This allows participants to penalize the model less severely when the visual cues are inherently subtle, ensuring a fairer and more stable evaluation of the model's intent-alignment capabilities.

To minimize cognitive load and ensure consistent judgments across samples, we developed a standardized annotation interface, as illustrated in Figure~\ref{fig:16}. For each sample, participants are presented with the video together with its generated audio and are asked to evaluate it along four predefined dimensions using a 5-point Likert scale. The interface also provides simple navigation controls, allowing participants to replay the sample and proceed through the evaluation in a structured and efficient manner.

\end{document}